\documentclass[journal]{IEEEtran}
\IEEEoverridecommandlockouts
\usepackage{setspace}
\usepackage{cite}
\usepackage{amsmath,amssymb,amsfonts,mathtools,bm}
\usepackage{amsthm}
\usepackage{algorithm}
\usepackage{algpseudocode}
\usepackage{graphicx}
\usepackage{textcomp}
\usepackage{xcolor,float}
\usepackage{tabularx}
\usepackage{textcomp}
\usepackage{diagbox}
\usepackage{svg}
\usepackage{booktabs}
\usepackage{subfigure}
\usepackage{hyperref}

\newtheorem{problem}{Problem}
\usepackage{graphicx}
\usepackage{makecell}
\usepackage{amsthm}
\usepackage{caption}
\usepackage{sidecap}
\usepackage{microtype}
\usepackage{booktabs} 
\usepackage{multirow}
\usepackage{float}
\usepackage{adjustbox} 
\usepackage{amsmath}
\usepackage{amssymb}
\usepackage{colortbl}
\usepackage{makecell}
\usepackage{float}
\usepackage{url}
\usepackage{balance}
\usepackage{bbding}
\usepackage{hyperref}
\usepackage{graphicx}
\usepackage{sidecap}
\usepackage{microtype}
\usepackage{graphicx}
\usepackage{subfigure}
\usepackage{booktabs} 
\usepackage{multirow}
\usepackage{array}
\usepackage{subcaption}
\usepackage{graphicx}
\usepackage{makecell}
\usepackage{amsmath}
\usepackage{amssymb}

\usepackage{tikz}
\usepackage{graphicx}
\usepackage{caption}
\usepackage{subcaption}

\usepackage{changes} 

\usepackage{booktabs}
\usepackage{multirow}
\usepackage{adjustbox}
\usepackage{xcolor}

\definecolor{ForestGreen}{rgb}{0.13,0.55,0.13}

\definecolor{lime}{HTML}{A6CE39}
\DeclareRobustCommand{\orcidicon}{%
    \begin{tikzpicture}
    \draw[lime, fill=lime] (0,0) 
    circle [radius=0.16] 
    node[white] {{\fontfamily{qag}\selectfont \tiny ID}};    \draw[white, fill=white] (-0.0625,0.095) 
    circle [radius=0.007];    \end{tikzpicture}
    \hspace{-2mm}}
\foreach \x in {A, ..., Z}{%
    \expandafter\xdef\csname orcid\x\endcsname{\noexpand\href{https://orcid.org/\csname orcidauthor\x\endcsname}{\noexpand\orcidicon}}
    }

\allowdisplaybreaks
\renewcommand{\baselinestretch}{1}

\begin{document}

\title{RadioDiff-Inv2: Differentiable Diffusion Inversion under Location Drift from Sparse Noisy Measurements for Radio Map Estimation}

\author{Xiucheng Wang, Kailong Wang, Nan Cheng

\thanks{ }
\thanks{
\par This work was supported by the National Key Research and Development Program of China (2020YFB1807700), and the National Natural Science Foundation of China (NSFC) under Grant No. 62201432.
\par Xiucheng Wang, Kailong Wang, and Nan Cheng are with the State Key Laboratory of ISN and School of Telecommunications Engineering, Xidian University, Xi’an 710071, China (e-mail: \{xcwang\_1, 24012100069\}@stu.xidian.edu.cn; dr.nan.cheng@ieee.org). 
\textit{Nan Cheng is the corresponding author}.
}

}

    \maketitle

\IEEEdisplaynontitleabstractindextext

\IEEEpeerreviewmaketitle

\begin{abstract}
Radio map (RM) estimation is a key enabler for environment-aware optimization in 6G wireless networks. In practice, RM construction increasingly relies on crowdsourced received signal strength (RSS) feedback that is inherently sparse and noisy. A further and often overlooked challenge is location drift, whereby privacy constraints and user mobility cause reported sampling coordinates to deviate from the true measurement locations. Unlike additive measurement noise, location drift perturbs the sensing operator itself, since each RSS sample effectively queries the underlying RM at an incorrect spatial coordinate. This operator uncertainty, compounded with sparse noisy sensing, renders the inverse problem severely ill-posed and limits conventional estimators that rely on analytically specified priors. This paper proposes RadioDiff-Inv2, a differentiable diffusion inversion framework that estimates RMs from sparse noisy measurements under location drift. A Gaussian resampling scheme is introduced to construct a differentiable, drift-aware measurement operator on grid-based maps, and the probability-flow ordinary differential equation (ODE) is exploited to cast the diffusion sampler as a deterministic, differentiable mapping from an initial noise code to the estimated RM. By optimizing the noise code via backpropagation against a drift-marginalized data-fidelity objective, RadioDiff-Inv2 produces reconstructions that are both prior-plausible and measurement-consistent without costly posterior sampling. Extensive experiments show that RadioDiff-Inv2 outperforms the best competing baseline by 4 to 14 dB in PSNR across varying sparsity and drift levels. The advantage is most pronounced in low-SNR regimes, where the learned diffusion prior maintains near-constant reconstruction fidelity while conventional methods degrade severely. Code is available at \url{https://github.com/UNIC-Lab/RadioDiff-inv2}.
\end{abstract}


\section{Introduction}

In 6G networks, environment-aware intelligence is expected to become a core capability, and radio maps (RMs) provide the spatially explicit propagation abstraction that underpins this vision \cite{wang2024tutorial,6g,shen2023toward}. By capturing how signal strength and shadowing vary across space, RMs directly support coverage planning \cite{zhang2020radio}, interference coordination \cite{wang2025radiodiff}, mobility-aware resource management \cite{raivio2003analysis}, and other proactive control functions that are difficult to realize from instantaneous link measurements alone \cite{zeng2021toward}. In operational settings, however, acquiring dense and reliable RMs at scale remains a major practical bottleneck \cite{wang2025radiodiff}. RM construction is increasingly driven by crowdsourced received signal strength (RSS) feedback collected opportunistically from user devices, because dedicated site surveys and dense infrastructure sensing are costly and hard to sustain \cite{yapar2025sampling,ye2018rmapcs}. While scalable, this crowdsourcing paradigm yields observations that are sparse in space and noisy in magnitude, owing to limited participation, heterogeneous hardware, and time-varying measurement conditions \cite{wang2024radiodiff}. Beyond sparsity and noise, crowdsourcing introduces a more fundamental uncertainty that reshapes the inverse problem, namely that the sampling locations themselves may be unreliable.

This location uncertainty, referred to as location drift, arises because privacy protection and user mobility prevent accurate localization of crowdsourced measurements \cite{yang2013freeloc}. In privacy-preserving deployments, fine-grained global positioning system (GPS) coordinates are often unavailable, obfuscated, or quantized before being reported to the network \cite{venkatraman2002location}, while in highly mobile scenarios measurements are collected on the move and are temporally misaligned with the assumed sampling time \cite{le2003mobile}. As a result, the reported coordinates can deviate from the true sampling positions by a random offset. Unlike additive amplitude noise, location drift perturbs the measurement operator itself by evaluating the latent RM at incorrect spatial coordinates \cite{chung2022improving}. This distinction is particularly consequential in 6G-relevant environments where propagation exhibits sharp spatial transitions near building edges, blockage boundaries, and regions shaped by diffraction and multipath \cite{deschamps1972ray}. Under such conditions, even moderate drift can map measurements across qualitatively different propagation regimes, amplifying reconstruction bias and rendering the inverse problem severely ill-posed \cite{slijepcevic2002location}. Consequently, the network observes only a sparse set of noisy RSS reports tied to imprecise coordinates, while the underlying RM must be inferred over the entire spatial domain, making both the field to be reconstructed and the effective sensing locations uncertain. The compound effect of location drift and low measurement SNR is especially detrimental, because high noise levels obscure the few spatial samples that are available while drift simultaneously misdirects the sensing operator, leaving conventional estimators with neither accurate values nor reliable locations to anchor the reconstruction.

Existing approaches to sparse RM reconstruction are fundamentally limited in addressing this compound challenge. Interpolation and kernel-based estimators, including Kriging-style methods \cite{sato2017kriging}, impose stationarity and smoothness assumptions that are frequently violated in realistic urban propagation environments where occlusions and diffraction induce sharp spatial transitions \cite{11278649,11083758}. When the measurement SNR is low, these estimators further amplify noise through the interpolation kernel, producing reconstructions whose quality degrades rapidly with decreasing SNR. Compressed sensing and maximum a posteriori (MAP) formulations combine a measurement fidelity term with a structural prior, yet their practical effectiveness depends critically on specifying a task-matched prior that faithfully captures the distribution of RMs across diverse scenes \cite{romero2024theoretical,wang2025radiodiff,sato2017kriging}. The difficulty of prior specification is further compounded by drift-induced sensing mismatch, where the likelihood itself becomes misspecified \cite{chung2023parallel}. Under low-SNR conditions, weak data-fidelity signals cannot adequately constrain the solution, and a misspecified prior leads to large estimation bias. Recent diffusion-based generative models have demonstrated strong capability in learning high-fidelity RM priors directly from data \cite{wang2024radiodiff,11278649,11282987}, yet translating these learned priors into measurement-consistent reconstructions under location drift remains nontrivial. Most measurement-guided diffusion formulations implicitly assume a fixed and correctly registered sensing operator \cite{wang2025radiodiffloc}, and under location drift this assumption breaks down, causing biased optimization signals and often necessitating expensive posterior sampling to maintain stability. These limitations are most acute in the low-SNR regime, where the measurements carry the least information and a strong, well-calibrated generative prior becomes indispensable for recovering the underlying field. These observations motivate a drift-aware inversion framework that can exploit diffusion-learned priors to provide robust reconstruction even when measurement quality is severely degraded.

To address these challenges, this paper proposes RadioDiff-Inv2, a differentiable diffusion inversion framework for RM estimation from sparse noisy measurements under location drift. The framework is built upon two coupled ideas. First, the probability-flow ordinary differential equation (ODE) \cite{chen2023probability} is exploited to cast the diffusion sampler as a deterministic, differentiable mapping from an initial noise code to the estimated RM, enabling gradient-based inversion by directly optimizing the noise code rather than relying on repeated stochastic sampling or costly posterior exploration. Second, a Gaussian resampling scheme is introduced to convert each drifted coordinate query into a smooth, differentiable measurement operator on grid-based maps, allowing gradients to propagate through both the diffusion ODE trajectory and the drift-aware sensing model. By optimizing the initial noise code to minimize a drift-marginalized data-fidelity objective via backpropagation, RadioDiff-Inv2 produces reconstructions that are simultaneously prior-plausible and consistent with sparse noisy drifted observations, while achieving substantially improved inference efficiency compared with sampling-based diffusion baselines. A distinctive advantage of this design is its robustness under low measurement SNR. Because the diffusion prior implicitly constrains the solution to lie on the learned RM manifold, the reconstruction quality is largely insensitive to noise severity, in contrast to conventional baselines whose performance degrades rapidly as the SNR decreases. The main contributions are summarized as follows.
\begin{enumerate}
    \item A drift-aware sparse inversion problem is formulated for 6G RM estimation from crowdsourced RSS, explicitly capturing the joint impact of measurement noise and location drift that renders the sensing operator uncertain.
    \item A Gaussian-resampling based differentiable drift-aware measurement operator is developed for grid-based RMs, enabling stable gradient propagation through the spatial misregistration induced by privacy constraints and user mobility.
    \item Sampling-free diffusion inversion is enabled by leveraging the probability-flow ODE to cast the diffusion generator as a deterministic, differentiable mapping from an initial noise code to the RM, and by optimizing this code to enforce drift-aware measurement consistency.
    \item RadioDiff-Inv2 is validated through extensive experiments, demonstrating improved reconstruction fidelity and robustness across a wide range of sparsity and drift levels, together with favorable inference efficiency over conventional baselines and sampling-based diffusion approaches. In particular, the method exhibits near-constant reconstruction quality across SNR levels from 0\,dB to 10\,dB, yielding over 15\,dB PSNR gain and more than 0.41 SSIM improvement over the best competing baseline at 0\,dB SNR, which confirms the strong regularization effect of the learned diffusion prior under extreme measurement corruption.
\end{enumerate}

\section{Preliminary and Related Works}
\subsection{Diffusion Models}
Denoising diffusion probabilistic models (DDPM) \cite{ho2020denoising}, define a generative model by progressively perturbing a clean sample $\bm{x}_0 \sim q(\bm{x}_0)$ into a sequence of latent variables $\{\bm{x}_t\}_{t=1}^T$ through Gaussian noise injection. Let $\{\beta_t\}_{t=1}^T$ denote a variance schedule with $\beta_t \in (0,1)$, and define $\alpha_t = 1-\beta_t$ and $\bar{\alpha}_t = \prod_{s=1}^t \alpha_s$. The forward diffusion is a Markov chain
\begin{align}
q(\bm{x}_t|\bm{x}_{t-1})=\mathcal{N}\!\left(\bm{x}_t;\sqrt{\alpha_t}\bm{x}_{t-1},(1-\alpha_t)\mathbf{I}\right).
\end{align}
By composition, the marginal distribution admits a closed form
\begin{align}
q(\bm{x}_t|\bm{x}_0)=\mathcal{N}\!\left(\bm{x}_t;\sqrt{\bar{\alpha}_t}\bm{x}_0,(1-\bar{\alpha}_t)\mathbf{I}\right),
\end{align}
which is equivalently represented as
\begin{align}
\bm{x}_t=\sqrt{\bar{\alpha}_t}\bm{x}_0+\sqrt{1-\bar{\alpha}_t}\bm{\epsilon},\qquad
\bm{\epsilon}\sim\mathcal{N}(\mathbf{0},\mathbf{I}).
\end{align}
DDPM constructs a reverse-time generative process $p_\theta(\bm{x}_{t-1}|\bm{x}_t)$ that aims to invert the forward corruption. The exact reverse posterior of the forward chain is also Gaussian:
\begin{align}
q(\bm{x}_{t-1}|\bm{x}_t,\bm{x}_0)=\mathcal{N}\!\left(\bm{x}_{t-1};\tilde{\bm{\mu}}(\bm{x}_t,\bm{x}_0,t),\tilde{\beta}_t\mathbf{I}\right),
\end{align}
with
\begin{align}
\tilde{\bm{\mu}}(\bm{x}_t,\bm{x}_0,t)
&=\frac{\sqrt{\alpha_t}(1-\bar{\alpha}_{t-1})}{1-\bar{\alpha}_t}\bm{x}_t
+\frac{\sqrt{\bar{\alpha}_{t-1}}(1-\alpha_t)}{1-\bar{\alpha}_t}\bm{x}_0,\\
\tilde{\beta}_t \label{eq:xt_eps_form}
&=\frac{1-\bar{\alpha}_{t-1}}{1-\bar{\alpha}_t}\beta_t.
\end{align}
In practice, the reverse transition is parameterized as
\begin{align}
p_\theta(\bm{x}_{t-1}|\bm{x}_t)=\mathcal{N}\!\left(\bm{x}_{t-1};\bm{\mu}_\theta(\bm{x}_t,t),\sigma_t^2\mathbf{I}\right),
\end{align}
where $\sigma_t^2$ is chosen as $\tilde{\beta}_t$ or a learnable variant, and $\bm{\mu}_\theta$ is produced by a neural network. A common and effective parameterization predicts the additive noise $\bm{\epsilon}$ in \eqref{eq:xt_eps_form}. Given a noise predictor $\bm{\epsilon}_\theta(\bm{x}_t,t)$, an estimate of the clean sample is obtained by
\begin{align}
\hat{\bm{x}}_0(\bm{x}_t,t)=\frac{\bm{x}_t-\sqrt{1-\bar{\alpha}_t}\,\bm{\epsilon}_\theta(\bm{x}_t,t)}{\sqrt{\bar{\alpha}_t}}.\label{eq:x0_hat}
\end{align}
Substituting $\hat{\bm{x}}_0$ into the posterior mean yields the standard DDPM sampling mean
\begin{align}
\bm{\mu}_\theta(\bm{x}_t,t)=\tilde{\bm{\mu}}\!\left(\bm{x}_t,\hat{\bm{x}}_0(\bm{x}_t,t),t\right).
\end{align}
Training is typically performed by uniformly sampling $t \in \{1,\ldots,T\}$ and minimizing the noise-prediction objective
\begin{align}
\min_\theta \ \mathbb{E}_{t,\bm{x}_0,\bm{\epsilon}}
\left[\left\|\bm{\epsilon}-\bm{\epsilon}_\theta(\bm{x}_t,t)\right\|_2^2\right],
\end{align}
This objective corresponds to a variational bound optimization under standard choices of $\sigma_t^2$, and it yields a model that can sample by iterating $p_\theta(\bm{x}_{t-1}|\bm{x}_t)$ from $\bm{x}_T \sim \mathcal{N}(\mathbf{0},\mathbf{I})$ down to $\bm{x}_0$.

Denoising diffusion implicit models (DDIM) \cite{song2020denoising}, retain the same forward marginals $q(\bm{x}_t|\bm{x}_0)$ and reuse the same trained predictor $\bm{\epsilon}_\theta(\bm{x}_t,t)$, but modify the reverse trajectory to a non-Markovian family that can be made deterministic. Given $\bm{x}_t$, DDIM first computes $\hat{\bm{x}}_0(\bm{x}_t,t)$ as in \eqref{eq:x0_hat} and then updates
\begin{align}
\bm{x}_{t-1}
=\sqrt{\bar{\alpha}_{t-1}}\hat{\bm{x}}_0(\bm{x}_t,t)
+\sqrt{1-\bar{\alpha}_{t-1}-\sigma_t^2}\,\bm{\epsilon}_\theta(\bm{x}_t,t)
+\sigma_t \bm{z},
\end{align}
where $\bm{z}\sim\mathcal{N}(\mathbf{0},\mathbf{I}),$, the noise scale $\sigma_t$ is controlled by a parameter $\eta \in [0,1]$ through
\begin{align}
\sigma_t
=\eta \sqrt{\frac{1-\bar{\alpha}_{t-1}}{1-\bar{\alpha}_t}}
\sqrt{1-\frac{\bar{\alpha}_t}{\bar{\alpha}_{t-1}}}.
\end{align}
Setting $\eta=1$ recovers a stochastic trajectory that matches the DDPM variance choice, while setting $\eta=0$ yields a fully deterministic sampler:
\begin{align}
\bm{x}_{t-1}
=\sqrt{\bar{\alpha}_{t-1}}\hat{\bm{x}}_0(\bm{x}_t,t)
+\sqrt{1-\bar{\alpha}_{t-1}}\,\bm{\epsilon}_\theta(\bm{x}_t,t).
\end{align}
A key implication is that, for $\eta=0$, the generated sample becomes a deterministic function of the initial noise $\bm{x}_T$, establishing a one-to-one mapping from the initial latent code to the reconstructed output. In addition, DDIM naturally supports accelerated sampling by operating on a subsequence of timesteps, since the update depends on $\bar{\alpha}_t$ rather than requiring all intermediate Markov transitions.

\subsection{Radio Map Estimation}
RM estimation from sparse measurements has been studied for years, starting from classical spatial interpolation and statistical learning \cite{wang2026tutorial}. Representative techniques include inverse distance weighting, Kriging \cite{sato2017kriging}, and Gaussian process regression \cite{zhang2022k}, which extrapolate a dense field from a limited set of samples by imposing stationarity and smoothness through hand-crafted kernels. While these methods are attractive for their simplicity and interpretability, they often struggle in outdoor urban environments where received power exhibits sharp transitions induced by blockage and diffraction. To reduce model mismatch and improve scalability, learning-based reconstruction has become a dominant paradigm \cite{chaves2023deeprem}. Convolutional encoder decoder frameworks such as RadioUNet \cite{levie2021radiounet} and its variants learn a direct mapping from sparse measurements and environment context to dense RMs, and deep denoising regularization has also been explored to improve robustness under measurement noise \cite{wang2026tutorial}. These point-estimation approaches can deliver strong accuracy under moderate sparsity, but they tend to oversmooth fine structures and their performance degrades when observations are extremely sparse or unevenly distributed.

To better capture long-range dependencies and global propagation patterns, recent works have introduced architectures beyond pure convolution. Conditional adversarial learning, including Pix2Pix-style formulations and task-driven variants such as TiRE-GAN \cite{zhou2025tire} and RME-GAN \cite{zhang2023rme}, can sharpen shadow boundaries and enhance high-frequency details, yet they may introduce visually plausible but physically inconsistent artifacts when the adversarial objective overwhelms data fidelity. Transformer-based models aim to model global interactions more explicitly. Designs such as RMTransformer \cite{li2025rmtransformer}, MARS \cite{deng2025mars}, and RadioFormer \cite{fang2025radioformer} leverage attention to aggregate sparse evidence across the map and improve reconstruction in complex layouts. Graph neural networks provide a complementary direction by representing measurement locations and prediction locations as nodes and propagating information via message passing, with RadioGAT \cite{li2024radiogat} as a representative example. These advances improve the ability to exploit irregular sampling patterns, but most formulations still assume that each measurement is correctly registered to its reported coordinate, so the sensing operator is treated as fixed during inference.

Generative modeling has recently emerged as a powerful tool to learn radio-map priors beyond deterministic regressors. Diffusion-based frameworks such as RadioDiff \cite{wang2024radiodiff} demonstrate strong capability in learning high-fidelity radio-map distributions, and physics-aware variants such as RMDM \cite{jia2025physics} and iRadioDiff \cite{wang2025iradiodiff} incorporate propagation-related constraints to enhance physical consistency. A practical concern for diffusion generation is inference latency due to iterative denoising, which has motivated acceleration efforts including flow-matching based models such as RadioFlow \cite{jia2025radioflow} that generate samples through an ordinary differential equation solver with fewer steps. For inverse problems, diffusion priors are commonly exploited through measurement guidance, plug-and-play style regularization, or posterior sampling \cite{wang2025radiodiff}. However, these strategies typically assume a correctly specified and registered sensing operator. Under spatial misregistration, naive guidance can yield biased optimization signals, while sampling-based posterior inference becomes increasingly expensive because the uncertainty in the sensing operator must be resolved jointly with the latent field.

This paper differs from existing sparse radio-map reconstruction and diffusion-based estimation in one key aspect: it explicitly addresses location drift in crowdsourced sensing, where privacy constraints and user mobility cause reported coordinates to deviate from true sampling locations. In this setting, the measurements are not only sparse and noisy, but also associated with an uncertain spatial query, which fundamentally changes the inference problem. By formulating drift-aware radio-map estimation with a drift-marginalized likelihood and by developing a differentiable inversion mechanism that enforces measurement consistency under coordinate uncertainty, the proposed framework targets a realism gap that is typically overlooked in prior learning-based and generative approaches.

\begin{figure*}
\captionsetup{font={small}, skip=16pt}
    \centering
    \includegraphics[width=1\linewidth]{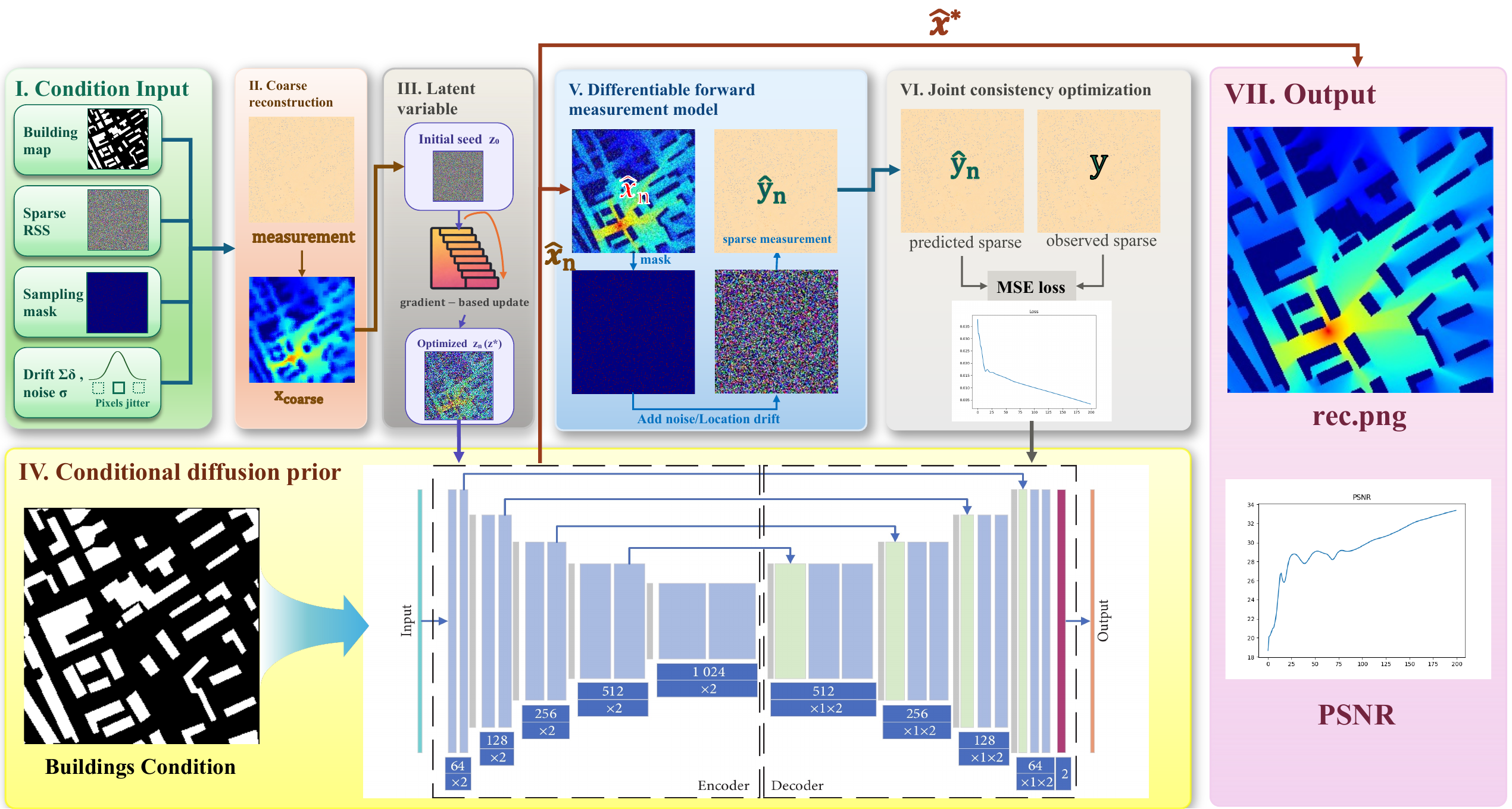}
    \caption{Overview of the proposed RadioDiff-Inv2 inference pipeline under sparse noisy measurements with Gaussian location drift. An initial noise seed serves as the optimization variable and is fed into a diffusion backbone conditioned on the environment occupancy map to generate a dense radio map. A differentiable post-process applies drift-aware Gaussian resampling and spatial masking to produce predicted sparse observations aligned with reported coordinates. The mean squared error between predicted and collected sparse RSS measurements is backpropagated to update the initial seed, yielding a reconstruction that is both prior-plausible and measurement-consistent under location uncertainty.}
    \label{fig-system}
\end{figure*}

\section{System Model and Problem Formulation}
\subsection{Environment Model}
In this paper, we consider an outdoor region of interest denoted by $\Omega \subset \mathbb{R}^2$. The region is discretized into a uniform grid with $N_x$ cells along the horizontal axis and $N_y$ cells along the vertical axis. Let $\mathcal{G}=\{\mathbf{p}_{i,j}\}$ denote the set of grid-center locations, where $\mathbf{p}_{i,j}\in\Omega$ is the physical coordinate associated with the cell indexed by $i$ and $j$. The grid resolution is characterized by the cell spacings $\Delta_x$ and $\Delta_y$, which determine the spatial granularity of the RM representation and the measurement operator used in the subsequent inversion. The distribution of buildings is represented by a binary occupancy matrix $\mathbf{H}\in\{0,1\}^{N_x\times N_y}$. An element $\mathbf{H}_{i,j}=1$ indicates that the corresponding cell is occupied by a building and is therefore treated as a blocked region for outdoor propagation. An element $\mathbf{H}_{i,j}=0$ indicates that the cell belongs to free space and can host valid receiver locations. The matrix $\mathbf{H}$ provides a compact geometric description of the environment and is used to determine visibility, blockage, and interaction events in the propagation model. The target quantity is a RM $\mathbf{R}\in\mathbb{R}^{N_x\times N_y}$ defined on the same grid. For each free-space cell, the entry $\mathbf{R}_{i,j}$ represents the large-scale received signal strength associated with location $\mathbf{p}_{i,j}$ under a given transmitter configuration and propagation environment. Throughout this paper, the RM values are modeled in a logarithmic power scale so that measurement noise and model mismatch can be handled consistently in the observation model and the drift-aware inversion procedure.

\subsection{Propagation Model}
Given the transmitter configuration and the environment map $\mathbf{H}$, the received signal strength at any receiver location $\mathbf{p}\in\Omega$ is predicted by a physics-based propagation operator that approximates electromagnetic wave propagation in outdoor scenes. The model is treated as a deterministic forward operator
\begin{align}
\mathrm{RSS}(\mathbf{p})=\mathcal{F}\!\left(\mathbf{p};\mathbf{H}\right),
\end{align}
which maps the receiver location and the environment to the corresponding received power level.

In logarithmic power scale, the received signal strength at location $\mathbf{p}$ is expressed by a link-budget form
\begin{align}
\mathrm{RSS}(\mathbf{p}) = P_{\mathrm{t}} + G_{\mathrm{t}}(\mathbf{p}) + G_{\mathrm{r}}(\mathbf{p}) - L(\mathbf{p}),
\end{align}
where $P_{\mathrm{t}}$ denotes the transmit power, $G_{\mathrm{t}}(\mathbf{p})$ denotes the transmit antenna gain along the effective propagation direction to $\mathbf{p}$, $G_{\mathrm{r}}(\mathbf{p})$ denotes the receive antenna gain, and $L(\mathbf{p})$ denotes the effective path loss in decibels.

The effective path loss is modeled by combining the free-space attenuation, the distance-dependent term, and additional interaction losses induced by the environment, written as
\begin{align}
L(\mathbf{p}) = &20\log_{10}\!\Big(\frac{4\pi}{\lambda}\Big)
+ 10\,\kappa(\mathbf{p})\log_{10}\!\big(d(\mathbf{p})\big)\notag\\
&+ L_{\mathrm{int}}(\mathbf{p})
- G_{\mathrm{wg}}(\mathbf{p})
- G_{\mathrm{t}}(\mathbf{p}),
\end{align}
where $\lambda$ is the carrier wavelength, $d(\mathbf{p})$ is the effective propagation path length associated with the dominant propagation route to $\mathbf{p}$, and $\kappa(\mathbf{p})$ is the scenario-dependent path loss exponent determined by the propagation condition at $\mathbf{p}$. The term $L_{\mathrm{int}}(\mathbf{p})$ aggregates interaction losses contributed by environment-induced mechanisms such as reflections and diffractions along the dominant route, and $G_{\mathrm{wg}}(\mathbf{p})$ captures potential waveguide-like gain effects in outdoor street canyons. The antenna gain $G_{\mathrm{t}}(\mathbf{p})$ is explicitly included to account for directional radiation and its impact on spatial RSS variations.

To characterize the accumulation of interaction losses, $L_{\mathrm{int}}(\mathbf{p})$ is expressed as a sum over interaction events along the dominant route
\begin{align}
L_{\mathrm{int}}(\mathbf{p})=\sum_{k=1}^{K(\mathbf{p})}\ell_k(\mathbf{p}),
\end{align}
where $K(\mathbf{p})$ denotes the number of interaction events and $\ell_k(\mathbf{p})$ denotes the loss associated with the $k$th event. The environment map $\mathbf{H}$ governs visibility and blockage, thereby determining whether a feasible route exists and shaping $d(\mathbf{p})$, $K(\mathbf{p})$, and the corresponding loss and gain terms, which in turn induces sharp spatial transitions in $\mathrm{RSS}(\mathbf{p})$ near occlusion boundaries in outdoor scenes.

\subsection{Measurement Model}
Crowdsourced sensing provides a set of spatially sparse received signal strength observations over the region of interest. Let $\mathcal{D}=\{(\tilde{\mathbf{p}}_m,y_m)\}_{m=1}^{M}$ denote the measurement set, where $\tilde{\mathbf{p}}_m\in\Omega$ is the reported coordinate of the $m$th sample and $y_m$ is the corresponding RSS value in a logarithmic power scale. The number of measurements is much smaller than the number of grid cells, so the sensing process is strongly under-determined. The reported RSS value is modeled as a noisy observation of the latent RM evaluated at the true sensing location. Specifically, for each measurement, the amplitude noise is modeled by an additive term
\begin{align}
y_m=\mathrm{RSS}(\mathbf{p}_m)+n_m,
\end{align}
where $\mathbf{p}_m$ is the unknown true sampling location and $n_m$ captures aggregate measurement uncertainty, including device heterogeneity, reporting errors, and residual small-scale fluctuations not represented by the propagation model. Due to privacy protection and user mobility, the true sampling location $\mathbf{p}_m$ is not directly available. Instead, the network only observes the reported coordinate $\tilde{\mathbf{p}}_m$, which deviates from $\mathbf{p}_m$ by a random drift. We model this location drift as a Gaussian perturbation
\begin{align}
\mathbf{p}_m=\tilde{\mathbf{p}}_m+\boldsymbol{\delta}_m,\qquad
\boldsymbol{\delta}_m\sim\mathcal{N}(\mathbf{0},\mathbf{\Sigma}_{\delta}),
\end{align}
where $\mathbf{\Sigma}_{\delta}$ controls the drift scale and anisotropy. This drift model implies that each RSS report is associated with an uncertain spatial query of the underlying RM. As a consequence, the sensing operator is not fixed by the reported coordinates alone, since the observation is generated at $\mathbf{p}_m$ rather than $\tilde{\mathbf{p}}_m$. This operator uncertainty, combined with sparse noisy sensing, significantly increases the ill-posedness of RM estimation and motivates drift-aware inference mechanisms in the subsequent sections.

\subsection{Problem Formulation}
Given the environment map $\mathbf{H}$, the transmitter configuration, the propagation operator $\mathcal{F}(\cdot;\mathbf{H})$, and a crowdsourced dataset $\mathcal{D}=\{(\tilde{\mathbf{p}}_m,y_m)\}_{m=1}^{M}$, the objective is to estimate a dense RM $\mathbf{R}\in\mathbb{R}^{N_x\times N_y}$ over the region of interest. The RM is defined on the grid $\mathcal{G}$, and $\mathbf{R}_{i,j}$ represents the received signal strength associated with the grid location $\mathbf{p}_{i,j}$. For cells occupied by buildings, indicated by $\mathbf{H}_{i,j}=1$, the RM values are not queried by valid user locations and are excluded by a spatial mask in the observation model. Each measurement $y_m$ is generated at an unknown true location $\mathbf{p}_m$ that deviates from the reported coordinate $\tilde{\mathbf{p}}_m$ by a Gaussian drift, and is further perturbed by additive measurement noise. Let $\mathcal{S}(\mathbf{R},\mathbf{p})$ denote the sampling operator that returns the RM value at a continuous coordinate $\mathbf{p}$ through a grid-based query rule. The measurement model is written as
\begin{align}
y_m &= \mathcal{S}(\mathbf{R},\mathbf{p}_m) + n_m, \qquad n_m \sim \mathcal{N}(0,\sigma_n^2), \label{eq:meas_model_form}\\
\mathbf{p}_m &= \tilde{\mathbf{p}}_m + \boldsymbol{\delta}_m, \qquad \boldsymbol{\delta}_m \sim \mathcal{N}(\mathbf{0},\mathbf{\Sigma}_{\delta}), \label{eq:drift_form}
\end{align}
where $\sigma_n^2$ and $\mathbf{\Sigma}_{\delta}$ are assumed known or estimated from system statistics. The inference target is the posterior distribution of $\mathbf{R}$ given the reported coordinates and noisy RSS measurements, after marginalizing the unknown drift variables. Let $p(\mathbf{R})$ denote a prior distribution over RMs. The drift-marginalized likelihood of each measurement is obtained by integrating over the Gaussian drift distribution. This leads to the following drift-aware maximum a posteriori formulation.
\begin{problem}\label{prob:drift_aware_rm}
\begin{align}
\max_{\mathbf{R}}&\;
 \log p(\mathbf{R})
+ \sum_{m=1}^{M} \log p\!\left(y_m \mid \mathbf{R}, \tilde{\mathbf{p}}_m\right), \label{eq:map_obj}\\
s.t.\quad & p\!\left(y_m \mid \mathbf{R}, \tilde{\mathbf{p}}_m\right)
=\int p\!\left(y_m \mid \mathbf{R}, \tilde{\mathbf{p}}_m+\boldsymbol{\delta}\right)
\, p(\boldsymbol{\delta})\, d\boldsymbol{\delta},\tag{\ref{eq:map_obj}a} \label{eq:marg_like}\\
&p\!\left(y_m \mid \mathbf{R}, \tilde{\mathbf{p}}_m+\boldsymbol{\delta}\right)
=\mathcal{N}\!\left(y_m;\mathcal{S}\!\left(\mathbf{R},\tilde{\mathbf{p}}_m+\boldsymbol{\delta}\right),\sigma_n^2\right),\tag{\ref{eq:map_obj}b} \label{eq:cond_like}\\
&p(\boldsymbol{\delta})
=\mathcal{N}\!\left(\boldsymbol{\delta};\mathbf{0},\mathbf{\Sigma}_{\delta}\right). \tag{\ref{eq:map_obj}c}\label{eq:drift_prior}
\end{align}
\end{problem}
\noindent Problem~\ref{prob:drift_aware_rm} is severely ill-posed because the number of measurements is far smaller than the number of grid values, while the sensing operator is uncertain due to the unknown drift. The next section develops an inference framework that leverages a learned generative prior and a differentiable drift-aware measurement mechanism to solve the above formulation efficiently.

\section{RadioDiff-Inv2 Framework}
\subsection{Prior Adaptation}
The proposed framework builds upon the denoising backbone of RadioDiff and preserves its core architectural principles. Specifically, we reuse the same multi-scale U-shaped denoiser with time-dependent embeddings and hierarchical feature fusion, so that the model retains the strong generative capacity learned from large-scale radio-map data. The forward diffusion schedule and the standard noise-prediction training objective are also kept unchanged, which ensures that the adapted model remains a valid diffusion prior over RMs within the considered environment class.

A key modification is made to the conditioning interface. RadioDiff is trained as a fully conditional diffusion model, where the denoiser is guided by two spatial condition channels that encode the environment occupancy and the transmitter location. In RM estimation from crowdsourced sensing, however, the transmitter location is often unavailable to the network due to limited metadata access and privacy constraints. Moreover, when both the environment and the transmitter location are known, the RM is largely determined by physics-based propagation modeling, and sparse RSS measurements provide limited additional information for recovering the field. This motivates an inference prior that does not rely on an explicit transmitter-location condition and can instead serve as an environment-specific prior for RMs.

To achieve transmitter-agnostic conditioning while keeping the network interface intact, we replace the transmitter-location condition channel with a replicated copy of the environment occupancy map. As a result, the denoiser still receives two spatial condition channels with the same dimensionality as in RadioDiff, but the conditioning semantics become environment-only. This design yields a diffusion prior that is specific to a given environment layout yet independent of transmitter-location inputs, and it is therefore compatible with the subsequent inversion stage where transmitter information is not assumed available.

The adaptation is performed with lightweight finetuning. The denoising backbone is kept fixed, and only the condition-related modules are updated, including the condition encoder and the layers that inject condition features into the denoiser. By limiting parameter updates to these components, the adapted model aligns its conditioning pathway with the new environment-only semantics while largely preserving the pretrained generative prior. This significantly reduces training cost and avoids retraining the diffusion model from scratch.

\subsection{Deterministic Mapping}
Let $\bm{R}\in\mathbb{R}^{N_x\times N_y}$ denote the target RM on the grid, and let $\mathbf{H}$ denote the environment occupancy map used as the condition after prior adaptation. We view reverse-time diffusion generation as a deterministic mapping from an initial seed to the final RM under a fixed condition. Specifically, consider a deterministic sampler with a decreasing time index set $\mathcal{T}=\{t_K>t_{K-1}>\cdots>t_1>t_0\}$, where $t_K$ corresponds to the noisiest state and $t_0$ corresponds to the final output. Given an initial seed $\bm{z}$ drawn from the standard Gaussian distribution, the sampler produces a sequence $\{\bm{x}_{t_k}\}_{k=0}^{K}$ by repeatedly applying a deterministic update rule
\begin{align}
\bm{x}_{t_{k-1}}=\Phi_{t_k\rightarrow t_{k-1}}\!\left(\bm{x}_{t_k};\mathbf{H}\right), \qquad k=1,2,\ldots,K,
\end{align}
where $\Phi_{t_k\rightarrow t_{k-1}}(\cdot)$ is defined by the denoiser prediction and the chosen deterministic trajectory. The overall reverse process is therefore a composition of deterministic transforms,
\begin{align}
\bm{R}=\mathcal{R}\!\left(\bm{z};\mathbf{H}\right)
=\Phi_{t_1\rightarrow t_0}\circ\Phi_{t_2\rightarrow t_1}\circ\cdots\circ\Phi_{t_K\rightarrow t_{K-1}}\!\left(\bm{z};\mathbf{H}\right),
\end{align}
which implies that, for a fixed environment condition $\mathbf{H}$ and fixed network parameters, the generated RM is uniquely determined by the seed $\bm{z}$.
\begin{algorithm}[t]
\caption{RadioDiff-Inv2 Inversion Under Gaussian Location Drift}
\label{alg:radiodiff_inv2}
\begin{algorithmic}[1]
\Require Environment map $\mathbf{H}$, measurements $\mathcal{D}=\{(\tilde{\mathbf{p}}_m,y_m)\}_{m=1}^{M}$, drift covariance $\mathbf{\Sigma}_{\delta}$, RSS noise variance $\sigma_n^2$
\Require Deterministic generator $\mathcal{R}(\cdot;\mathbf{H})$, resampling operator $\mathcal{T}_{\boldsymbol{\delta}}(\cdot)$, sampling operator $\mathcal{S}(\cdot,\cdot)$
\Require Seed update steps $K$, drift samples $S$, step sizes $\{\eta_k\}_{k=0}^{K-1}$
\Ensure Estimated RM $\hat{\mathbf{R}}$

\State Initialize seed $\mathbf{z}^{(0)} \sim \mathcal{N}(\mathbf{0},\mathbf{I})$
\For{$k=0$ \textbf{to} $K-1$}
    \State $\mathbf{R}^{(k)} \gets \mathcal{R}\!\left(\mathbf{z}^{(k)};\mathbf{H}\right)$
    \State $\mathcal{L} \gets 0$
    \For{$m=1$ \textbf{to} $M$}
        \State $\ell_m \gets 0$
        \For{$s=1$ \textbf{to} $S$}
            \State Sample drift $\boldsymbol{\delta}_{m}^{(s)} \sim \mathcal{N}(\mathbf{0},\mathbf{\Sigma}_{\delta})$
            \State $\mathbf{R}_{m}^{(k,s)} \gets \mathcal{T}_{\boldsymbol{\delta}_{m}^{(s)}}\!\left(\mathbf{R}^{(k)}\right)$
            \State $\hat{y}_{m}^{(k,s)} \gets \mathcal{S}\!\left(\mathbf{R}_{m}^{(k,s)},\tilde{\mathbf{p}}_m\right)$ with free-space masking from $\mathbf{H}$
            \State $\ell_m \gets \ell_m + \frac{1}{2\sigma_n^2}\left(y_m-\hat{y}_{m}^{(k,s)}\right)^2$
        \EndFor
        \State $\mathcal{L} \gets \mathcal{L} + \frac{1}{S}\ell_m$
    \EndFor
    \State $\mathbf{g}^{(k)} \gets \nabla_{\mathbf{z}} \mathcal{L}$ via backpropagation through $\mathcal{S}$, $\mathcal{T}_{\boldsymbol{\delta}}$, and $\mathcal{R}$
    \State $\mathbf{z}^{(k+1)} \gets \mathbf{z}^{(k)} - \eta_k \mathbf{g}^{(k)}$
    \If{a stopping criterion is met}
        \State \textbf{break}
    \EndIf
\EndFor
\State $\hat{\mathbf{R}} \gets \mathcal{R}\!\left(\mathbf{z}^{(k+1)};\mathbf{H}\right)$
\State \Return $\hat{\mathbf{R}}$
\end{algorithmic}
\end{algorithm}
A concrete instantiation is obtained by using a deterministic ordinary differential equation trajectory, which yields a sampling path without stepwise random noise injection. In this case, the reverse-time evolution is governed by an ODE whose drift is parameterized by the learned score or noise predictor, and the final map is given by the ODE solution at terminal time. Alternatively, a deterministic discrete-time sampler can be used, where each update computes an estimate of the clean sample from the current noisy state and then advances deterministically to the next time level. Both choices lead to the same essential property required by our inversion, namely that the mapping $\bm{z}\mapsto\mathcal{R}(\bm{z};\mathbf{H})$ is deterministic and differentiable almost everywhere with respect to the seed.

This deterministic mapping enables a seed-space reparameterization of the inverse problem. Instead of directly optimizing over the high-dimensional RM $\bm{R}$ under a hand-crafted regularizer, we restrict the search to the range of the generative mapping by writing $\bm{R}=\mathcal{R}(\bm{z};\mathbf{H})$ and optimizing the seed variable:
\begin{align}
\hat{\bm{z}}
=\arg\min_{\bm{z}}
\ \mathcal{L}_{\mathrm{data}}\!\left(\mathcal{M}\!\left(\mathcal{R}(\bm{z};\mathbf{H})\right), \bm{y}\right),
\qquad
\hat{\bm{R}}=\mathcal{R}\!\left(\hat{\bm{z}};\mathbf{H}\right),
\end{align}
where $\bm{y}$ denotes the sparse noisy measurements and $\mathcal{M}(\cdot)$ denotes the forward measurement operator that will be specified in the drift-aware setting. Since the diffusion prior is embedded in the generator range, the resulting estimate is encouraged to remain prior-plausible while the objective enforces measurement consistency. An additional benefit of seed-space optimization is that it avoids stepwise interventions that alternate between denoising dynamics and external correction operations. The reverse sampler is kept intact and is treated as a fixed differentiable function, while the measurement constraints are enforced globally through the optimization objective. This separation reduces the risk that local corrections at intermediate timesteps disrupt the generative trajectory, and it provides a stable gradient pathway from the measurement loss to the seed through the deterministic mapping $\mathcal{R}(\cdot;\mathbf{H})$.

\begin{figure*}
\centering
\setlength{\tabcolsep}{2pt}
\renewcommand{\arraystretch}{1.0}

\begin{tabular}{ccccccc}

\includegraphics[width=0.13\textwidth]{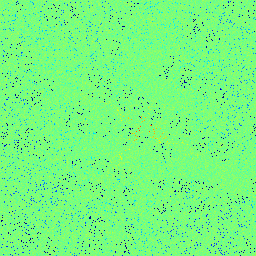} &
\includegraphics[width=0.13\textwidth]{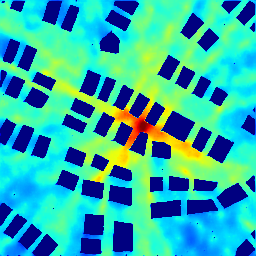} &
\includegraphics[width=0.13\textwidth]{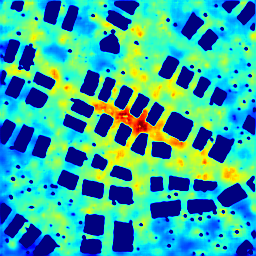} &
\includegraphics[width=0.13\textwidth]{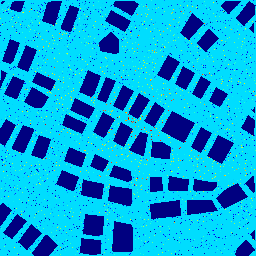} &
\includegraphics[width=0.13\textwidth]{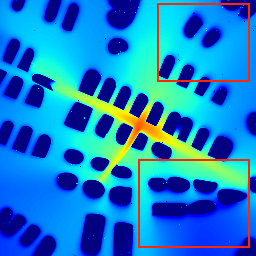} &
\includegraphics[width=0.13\textwidth]{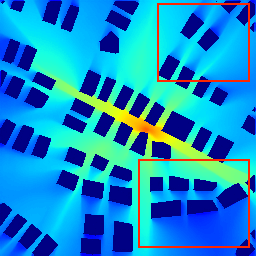} &
\includegraphics[width=0.13\textwidth]{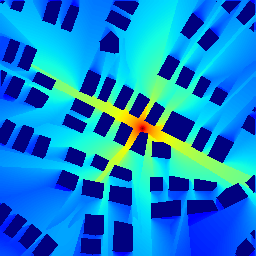} \\
{\scriptsize Samples} & {\scriptsize RME-GAN} & {\scriptsize RadioUNet} & {\scriptsize Kriging} & {\scriptsize RadioDiff-Inv1} & {\scriptsize RadioDiff-Inv2 (Ours)} & {\scriptsize Ground Truth} \\

\includegraphics[width=0.13\textwidth]{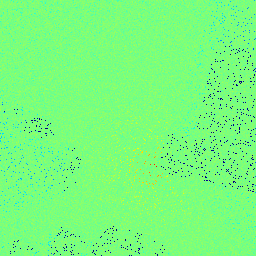} &
\includegraphics[width=0.13\textwidth]{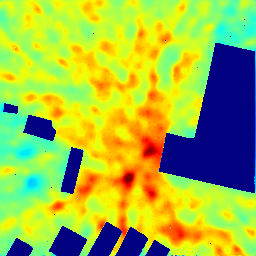} &
\includegraphics[width=0.13\textwidth]{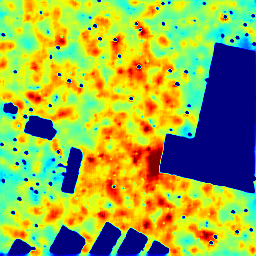} &
\includegraphics[width=0.13\textwidth]{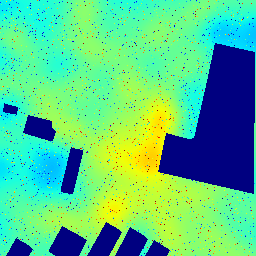} &
\includegraphics[width=0.13\textwidth]{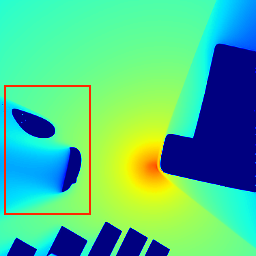} &
\includegraphics[width=0.13\textwidth]{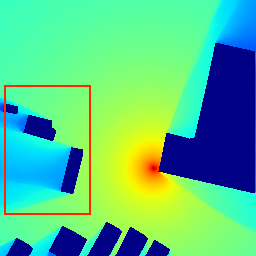} &
\includegraphics[width=0.13\textwidth]{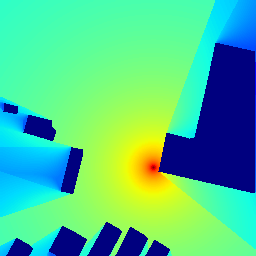} \\
{\scriptsize Samples} & {\scriptsize RME-GAN} & {\scriptsize RadioUNet} & {\scriptsize Kriging} & {\scriptsize RadioDiff-Inv1} & {\scriptsize RadioDiff-Inv2 (Ours)} & {\scriptsize Ground Truth} \\

\includegraphics[width=0.13\textwidth]{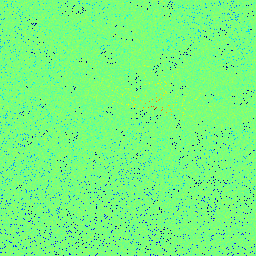} &
\includegraphics[width=0.13\textwidth]{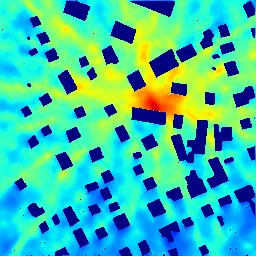} &
\includegraphics[width=0.13\textwidth]{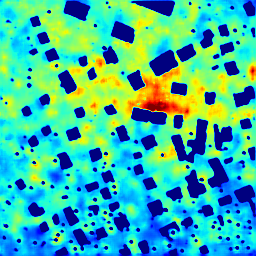} &
\includegraphics[width=0.13\textwidth]{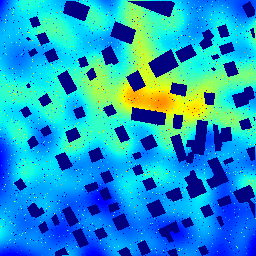} &
\includegraphics[width=0.13\textwidth]{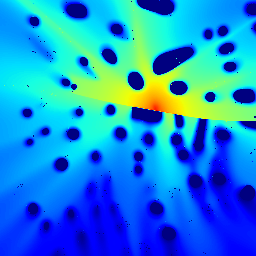} &
\includegraphics[width=0.13\textwidth]{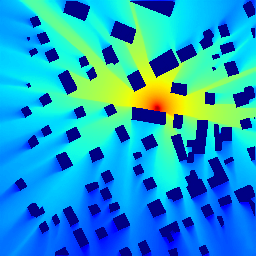} &
\includegraphics[width=0.13\textwidth]{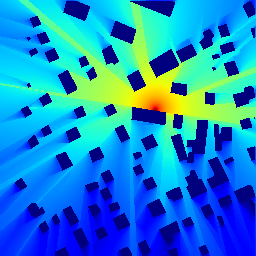} \\
{\scriptsize Samples} & {\scriptsize RME-GAN} & {\scriptsize RadioUNet} & {\scriptsize Kriging} & {\scriptsize RadioDiff-Inv1} & {\scriptsize RadioDiff-Inv2 (Ours)} & {\scriptsize Ground Truth} \\

\includegraphics[width=0.13\textwidth]{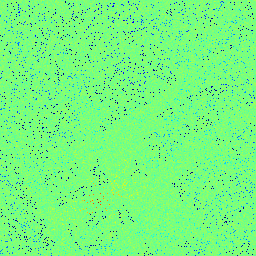} &
\includegraphics[width=0.13\textwidth]{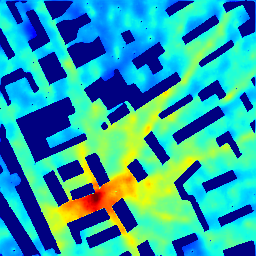} &
\includegraphics[width=0.13\textwidth]{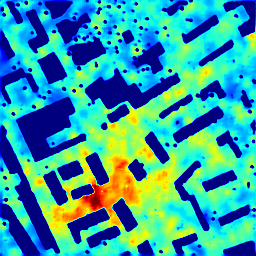} &
\includegraphics[width=0.13\textwidth]{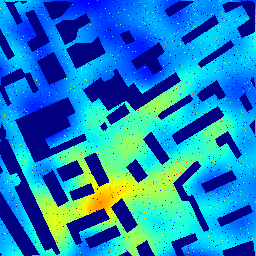} &
\includegraphics[width=0.13\textwidth]{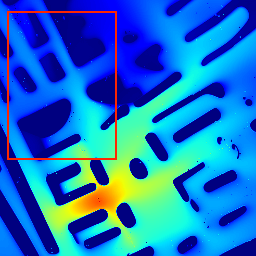} &
\includegraphics[width=0.13\textwidth]{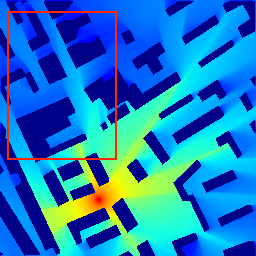} &
\includegraphics[width=0.13\textwidth]{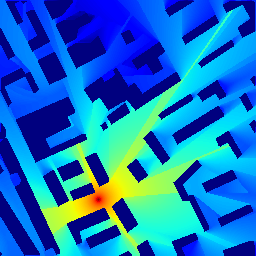} \\
{\scriptsize Samples} & {\scriptsize RME-GAN} & {\scriptsize RadioUNet} & {\scriptsize Kriging} & {\scriptsize RadioDiff-Inv1} & {\scriptsize RadioDiff-Inv2 (Ours)} & {\scriptsize Ground Truth} \\

\end{tabular}

\caption{Visual comparison under \textbf{Sampling Rate} $kp=0.05$, SNR=5, and $\Sigma_{\delta}=0.8$ (4 random samples).}
\label{fig:vis_kp005}
\end{figure*}

\subsection{Gaussian Resampling}
Location drift implies that each reported coordinate is only an imprecise proxy of the true sensing location. Let $\tilde{\mathbf{p}}_m$ denote the reported coordinate of the $m$th RSS report and let $\mathbf{p}_m$ denote the unknown true sensing location. Under the Gaussian drift model, $\mathbf{p}_m$ is distributed around $\tilde{\mathbf{p}}_m$ with a covariance determined by the drift statistics. Therefore, measurement consistency cannot be enforced by simply querying the reconstructed RM at $\tilde{\mathbf{p}}_m$. Instead, the observation should be consistent with the RM evaluated at uncertain spatial queries induced by drift. This motivates a drift-aware measurement operator that is differentiable with respect to the reconstructed RM and can be composed with the deterministic diffusion mapping for end-to-end inversion.

Let $\mathbf{R}$ be a grid-based RM and let $\mathcal{S}(\mathbf{R},\mathbf{p})$ denote the sampling operator that returns the RM value at a continuous coordinate $\mathbf{p}$ through bilinear interpolation on the grid. For a given drift realization $\boldsymbol{\delta}$, the predicted measurement associated with $\tilde{\mathbf{p}}_m$ is $\mathcal{S}(\mathbf{R},\tilde{\mathbf{p}}_m+\boldsymbol{\delta})$. Drift-aware consistency is obtained by marginalizing over the drift distribution, which leads to a drift-marginalized likelihood for each report. Since the marginalization does not admit a closed form for a general RM, we approximate it by Monte Carlo sampling. At each inversion iteration, we draw $S$ independent drift samples $\{\boldsymbol{\delta}_{m}^{(s)}\}_{s=1}^{S}$ from the Gaussian drift distribution and form the drift-aware data term by averaging the per-sample negative log-likelihood:
\begin{align}
\mathcal{L}_{\mathrm{drift}}(\mathbf{R})
=\sum_{m=1}^{M}\frac{1}{S}\sum_{s=1}^{S}
\ell\!\left(y_m,\mathcal{S}\!\left(\mathbf{R},\tilde{\mathbf{p}}_m+\boldsymbol{\delta}_{m}^{(s)}\right)\right),
\end{align}
where $\ell(\cdot,\cdot)$ is chosen according to the RSS noise model, such as a weighted squared error consistent with Gaussian amplitude noise. This construction yields an unbiased gradient estimator with respect to $\mathbf{R}$ when the sampling operator is differentiable, enabling backpropagation through the drift-aware data term.

To implement $\mathcal{S}(\mathbf{R},\tilde{\mathbf{p}}_m+\boldsymbol{\delta})$ efficiently on a grid, we use a resampling view. Evaluating $\mathbf{R}$ at a drifted coordinate is equivalent to applying a spatial shift to the entire map and then sampling at the reported coordinate. We therefore define a differentiable resampling operator $\mathcal{T}_{\boldsymbol{\delta}}(\cdot)$ that produces a shifted map $\mathbf{R}_{\boldsymbol{\delta}}=\mathcal{T}_{\boldsymbol{\delta}}(\mathbf{R})$ using bilinear interpolation. The predicted measurement can then be written as $\mathcal{S}(\mathbf{R}_{\boldsymbol{\delta}},\tilde{\mathbf{p}}_m)$. In practice, the resampling is implemented by grid-based bilinear interpolation, which is linear with respect to the input RM for a fixed shift. As a result, the gradient of the loss with respect to $\mathbf{R}$ is distributed back to the four neighboring pixels of each sampled location according to the same bilinear weights. This property provides stable gradients and avoids non-differentiable index operations during inversion.

The drift-aware operator is further coupled with the environment occupancy map $\mathbf{H}$ to ensure physical feasibility. We restrict valid sensing locations to free-space cells and exclude building-occupied cells from contributing to predicted measurements. This is achieved by applying a spatial mask derived from $\mathbf{H}$ during sampling and by renormalizing the interpolation weights when a neighborhood overlaps blocked cells, so that the operator remains numerically consistent and the predicted RSS is not contaminated by invalid regions. With this drift-aware differentiable measurement operator, the inversion objective can be optimized through the deterministic diffusion mapping by backpropagation, thereby enforcing measurement consistency under sparse sensing and Gaussian location drift.

\subsection{Inversion Algorithm}
We now present the end-to-end differentiable inversion procedure that combines the transmitter-agnostic diffusion prior, the deterministic reverse mapping, and the drift-aware Gaussian resampling operator. The key reparameterization is to express the unknown RM as the output of a deterministic generator driven by a seed variable. Let $\bm{z}$ denote the initial seed and let $\mathcal{R}(\bm{z};\mathbf{H})$ denote the deterministic reverse mapping under the environment condition $\mathbf{H}$. The estimated RM is written as
\begin{align}
\bm{R}=\mathcal{R}(\bm{z};\mathbf{H}). \label{eq:seed_param}
\end{align}
This parameterization restricts the search space to RMs that are consistent with the learned diffusion prior, while leaving the seed $\bm{z}$ as the optimization variable that can be adapted to match the measurements.

Let $\mathcal{D}=\{(\tilde{\mathbf{p}}_m,y_m)\}_{m=1}^{M}$ be the sparse crowdsourced dataset and let $\sigma_n^2$ denote the RSS noise variance. Drift-aware measurement consistency is enforced by Monte Carlo marginalization over Gaussian drift. At each iteration, we draw $S$ independent drift samples $\boldsymbol{\delta}_{m}^{(s)}\sim\mathcal{N}(\mathbf{0},\mathbf{\Sigma}_{\delta})$ and define the drift-aware data fidelity term as
\begin{align}
\mathcal{L}_{\mathrm{data}}(\bm{z})
=\sum_{m=1}^{M}\sum_{s=1}^{S}
\frac{1}{2S\sigma_n^2}
\left(y_m-\mathcal{S}\!\left(\mathcal{T}_{\boldsymbol{\delta}_{m}^{(s)}}\!\left(\mathcal{R}(\bm{z};\mathbf{H})\right),\tilde{\mathbf{p}}_m\right)\right)^2, \label{eq:data_loss_seed}
\end{align}
where $\mathcal{T}_{\boldsymbol{\delta}}(\cdot)$ denotes the bilinear resampling operator that shifts the RM according to the drift realization, and $\mathcal{S}(\cdot,\tilde{\mathbf{p}}_m)$ denotes the grid-based sampling operator that returns the predicted RSS at the reported coordinate. The environment occupancy map $\mathbf{H}$ is used to mask invalid regions during sampling so that measurements are matched only to free-space locations.

The inversion is performed by minimizing \eqref{eq:data_loss_seed} with respect to the seed variable:
\begin{align}
\hat{\bm{z}}=\arg\min_{\bm{z}} \ \mathcal{L}_{\mathrm{data}}(\bm{z}), 
\qquad 
\hat{\bm{R}}=\mathcal{R}(\hat{\bm{z}};\mathbf{H}). \label{eq:seed_opt}
\end{align}
The gradient of $\mathcal{L}_{\mathrm{data}}$ with respect to $\bm{z}$ is obtained by backpropagation through the complete computational graph. In particular, the gradient flows from the measurement residuals through the sampling operator, through the bilinear resampling operator, and through the deterministic reverse mapping implemented by the diffusion sampler. This end-to-end differentiability is essential for enforcing drift-aware consistency without modifying the reverse dynamics itself.

Algorithmically, the procedure iterates between deterministic generation and seed update. Starting from an initialization $\bm{z}^{(0)}\sim\mathcal{N}(\mathbf{0},\mathbf{I})$, each iteration generates a RM $\bm{R}^{(k)}=\mathcal{R}(\bm{z}^{(k)};\mathbf{H})$, evaluates the drift-aware data fidelity term using $S$ drift samples, and updates the seed by a first-order method
\begin{align}
\bm{z}^{(k+1)}=\bm{z}^{(k)}-\eta_k \nabla_{\bm{z}} \mathcal{L}_{\mathrm{data}}\!\left(\bm{z}^{(k)}\right), \label{eq:seed_update}
\end{align}
where $\eta_k$ is the step size. The algorithm terminates when the loss improvement becomes negligible or a maximum number of iterations is reached, and outputs $\hat{\bm{R}}$. This inference strategy offers two practical advantages. First, it avoids retraining a diffusion model for each sensing configuration. Training a diffusion model conditioned on sparse measurement maps ties the model to a particular distribution of sampling density, sampling geometry, noise level, and drift scale. When these factors change, the condition input can deviate substantially from the training distribution, leading to unstable generalization. In contrast, the proposed inversion keeps the diffusion prior fixed and absorbs the sensing variations into the drift-aware data term and operator, improving adaptability across heterogeneous crowdsourcing regimes. Second, the inference cost can be controlled by two knobs, namely the number of deterministic sampling steps used to evaluate $\mathcal{R}(\cdot;\mathbf{H})$ and the number of drift samples $S$ used in \eqref{eq:data_loss_seed}. Reducing either quantity lowers latency, while the resulting accuracy tradeoff can be characterized empirically in the evaluation section.

\begin{figure*}
\centering
\setlength{\tabcolsep}{2pt}
\renewcommand{\arraystretch}{1.0}

\begin{tabular}{ccccccc}

\includegraphics[width=0.13\textwidth]{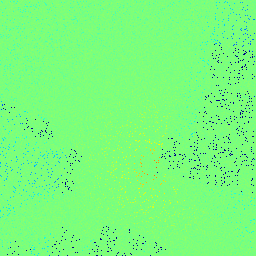} &
\includegraphics[width=0.13\textwidth]{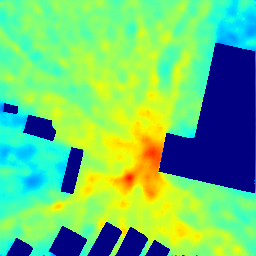} &
\includegraphics[width=0.13\textwidth]{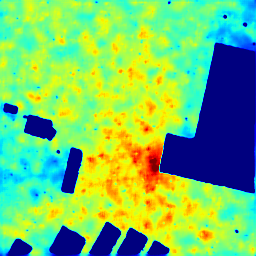} &
\includegraphics[width=0.13\textwidth]{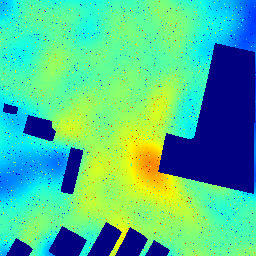} &
\includegraphics[width=0.13\textwidth]{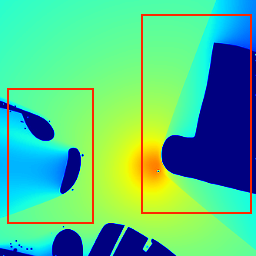} &
\includegraphics[width=0.13\textwidth]{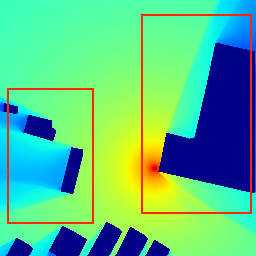} &
\includegraphics[width=0.13\textwidth]{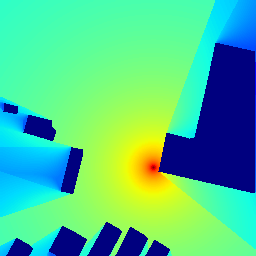} \\
{\scriptsize Samples} & {\scriptsize RME-GAN} & {\scriptsize RadioUNet} & {\scriptsize Kriging} & {\scriptsize RadioDiff-Inv1} & {\scriptsize RadioDiff-Inv2 (Ours)} & {\scriptsize Ground Truth} \\

\includegraphics[width=0.13\textwidth]{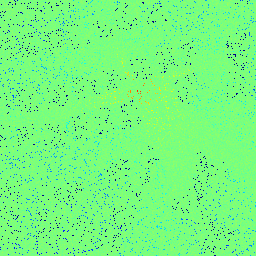} &
\includegraphics[width=0.13\textwidth]{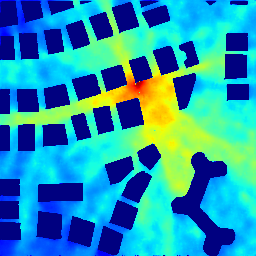} &
\includegraphics[width=0.13\textwidth]{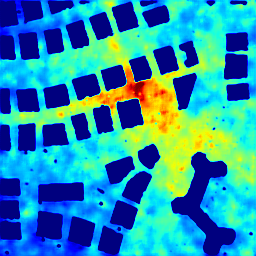} &
\includegraphics[width=0.13\textwidth]{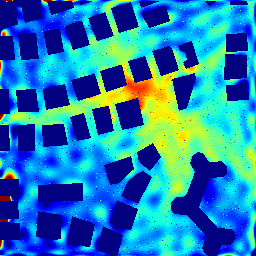} &
\includegraphics[width=0.13\textwidth]{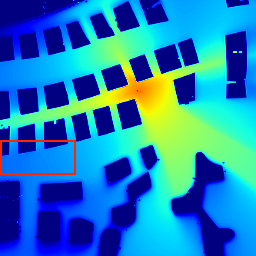} &
\includegraphics[width=0.13\textwidth]{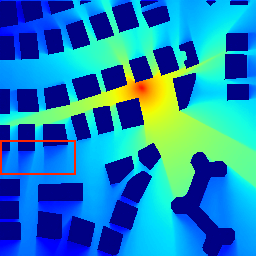} &
\includegraphics[width=0.13\textwidth]{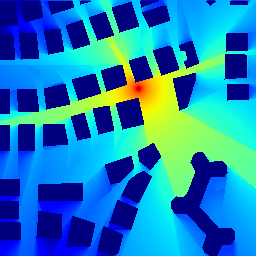} \\
{\scriptsize Samples} & {\scriptsize RME-GAN} & {\scriptsize RadioUNet} & {\scriptsize Kriging} & {\scriptsize RadioDiff-Inv1} & {\scriptsize RadioDiff-Inv2 (Ours)} & {\scriptsize Ground Truth} \\

\includegraphics[width=0.13\textwidth]{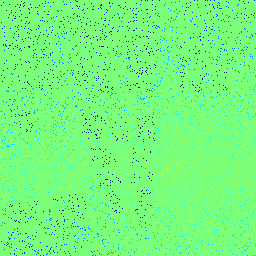} &
\includegraphics[width=0.13\textwidth]{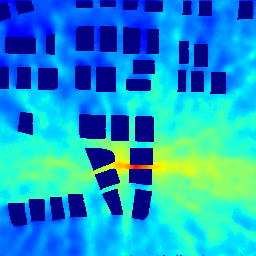} &
\includegraphics[width=0.13\textwidth]{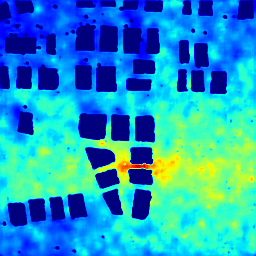} &
\includegraphics[width=0.13\textwidth]{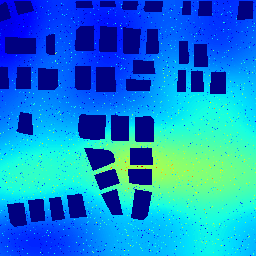} &
\includegraphics[width=0.13\textwidth]{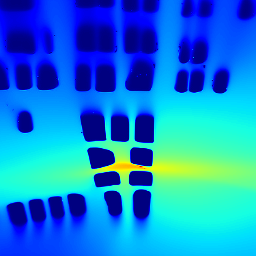} &
\includegraphics[width=0.13\textwidth]{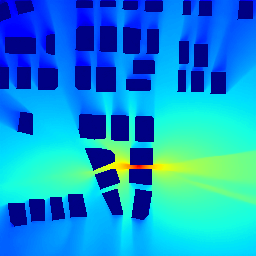} &
\includegraphics[width=0.13\textwidth]{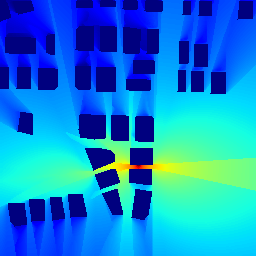} \\
{\scriptsize Samples} & {\scriptsize RME-GAN} & {\scriptsize RadioUNet} & {\scriptsize Kriging} & {\scriptsize RadioDiff-Inv1} & {\scriptsize RadioDiff-Inv2 (Ours)} & {\scriptsize Ground Truth} \\

\includegraphics[width=0.13\textwidth]{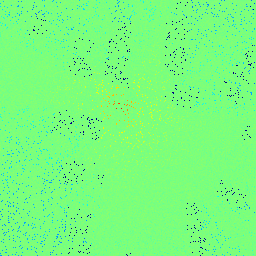} &
\includegraphics[width=0.13\textwidth]{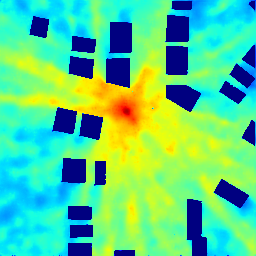} &
\includegraphics[width=0.13\textwidth]{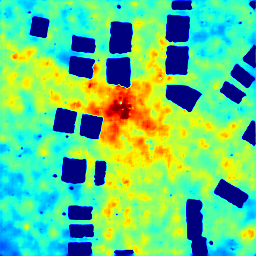} &
\includegraphics[width=0.13\textwidth]{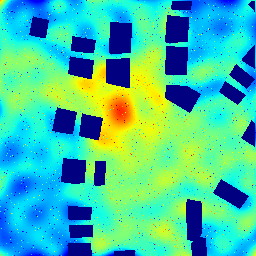} &
\includegraphics[width=0.13\textwidth]{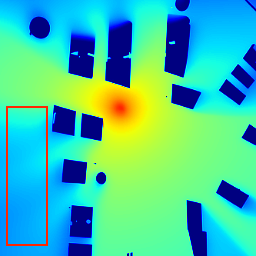} &
\includegraphics[width=0.13\textwidth]{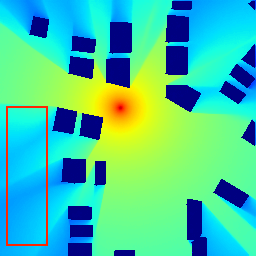} &
\includegraphics[width=0.13\textwidth]{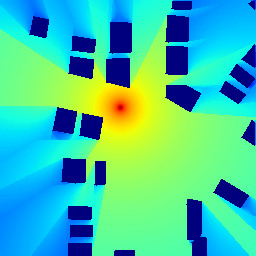} \\
{\scriptsize Samples} & {\scriptsize RME-GAN} & {\scriptsize RadioUNet} & {\scriptsize Kriging} & {\scriptsize RadioDiff-Inv1} & {\scriptsize RadioDiff-Inv2 (Ours)} & {\scriptsize Ground Truth} \\

\end{tabular}

\caption{Visual comparison under \textbf{SNR = 10}, $kp=0.05$, and $\Sigma_{\delta}=0.8$.}
\label{fig:vis_snr10}
\end{figure*}

\section{Experiment Results}

\subsection{Experimental Setup}
We conduct experiments on the RadioMapSeer dataset \cite{levie2021radiounet}, which comprises $701$ distinct urban environment topologies with varying building layouts. Among these, $600$ environments are used to train the diffusion prior, and the remaining $101$ environments are held out exclusively for testing. Each environment is represented by a binary occupancy map $\mathbf{H} \in \{0,1\}^{N_x \times N_y}$ and a corresponding set of ground-truth radio maps generated by a physics-based propagation simulator. This dataset provides a diverse and realistic benchmark for evaluating RM estimation across heterogeneous propagation conditions.

We compare RadioDiff-Inv2 against three representative baselines that span classical, learning-based, and generative estimation paradigms:
\begin{itemize}
    \item \textbf{Kriging} \cite{wang2026tutorial}: A classical spatial interpolation method based on Gaussian process regression that imposes stationarity and smoothness through a hand-crafted covariance kernel. It serves as a representative of non-parametric statistical estimators.
    \item \textbf{RadioUNet} \cite{levie2021radiounet}: A convolutional encoder-decoder network that learns a direct mapping from sparse measurements and environment context to dense RMs. It represents the supervised learning-based reconstruction paradigm.
    \item \textbf{RME-GAN} \cite{zhang2023rme}: A conditional generative adversarial network for RM estimation that sharpens shadow boundaries and high-frequency detail through adversarial training. It represents the GAN-based generative estimation paradigm.
    \item \textbf{RadioDiff-Inv1} \cite{wang2025radiodiff}: A diffusion-enhanced Bayesian inverse estimation framework that uses an unconditional generative diffusion model.
\end{itemize}

Four complementary metrics are adopted to provide a comprehensive assessment of reconstruction fidelity.
\begin{itemize}
    \item \textbf{PSNR} (Peak Signal-to-Noise Ratio, dB): measures pixel-level reconstruction accuracy; higher is better.
    \item \textbf{SSIM} (Structural Similarity Index): evaluates perceptual structural similarity; higher is better.
    \item \textbf{LPIPS} (Learned Perceptual Image Patch Similarity): captures high-level perceptual quality via deep features; lower is better.
    \item \textbf{NMSE} (Normalized Mean Squared Error): quantifies relative reconstruction error in signal power; lower is better.
\end{itemize}

Unless otherwise specified, experiments adopt a default sampling rate of $kp = 0.05$, a signal-to-noise ratio of $\mathrm{SNR} = 5\,\mathrm{dB}$, and a Gaussian location drift with covariance $\Sigma_\delta = 0.8$. Three groups of controlled experiments are conducted by varying one factor at a time: (i) sampling rate $kp \in \{0.005, 0.01, 0.05, 0.1\}$; (ii) measurement SNR $\in \{0, 1, 5, 10\}\,\mathrm{dB}$; and (iii) drift scale $\Sigma_\delta \in \{0.0, 0.8, 1.6, 2.4, 3.2\}$.

\newcommand{\diffgoodup}[1]{\textcolor{ForestGreen}{(#1)}} 
\newcommand{\diffgooddown}[1]{\textcolor{blue}{(#1)}}      
\newcommand{\diffbad}[1]{\textcolor{red}{(#1)}}            
\newcommand{\diffsecond}[1]{\textcolor{gray}{(#1)}}        

\begin{table*}[!t]
\centering
\begin{minipage}{0.49\textwidth}
\centering
\caption{PSNR comparison under different \textbf{Sampling Rate}, \textbf{SNR}, and \textbf{$\Sigma_{\delta}$} settings (higher is better). The best result in each row is highlighted in bold, the second-best is underlined, and the relative gain is computed as $(\text{best}-\text{second})/\text{second}$.}
\label{tab:psnr_all}
\begin{adjustbox}{max width=\textwidth}
\begin{tabular}{c|ccccc}
\toprule
Setting & RadioDiff-Inv2 (Ours) & RadioDiff-Inv1 & Kriging & RadioUNet & RME-GAN \\
\midrule
\multicolumn{6}{c}{\textbf{Sampling Rate (SNR=5, $\Sigma_{\delta}$=0.8)}}\\
\midrule
kp=0.005 & \textbf{23.8166} \textcolor{ForestGreen}{(+18.92\%)} & 16.4509 & \underline{20.0272} & 18.4192 & 18.6240 \\
kp=0.01  & \textbf{25.9965} \textcolor{ForestGreen}{(+31.46\%)} & 17.5264 & \underline{19.7745} & 18.3819 & 18.0740 \\
kp=0.05  & \textbf{31.3158} \textcolor{ForestGreen}{(+33.38\%)} & \underline{23.4786} & 21.3934 & 16.5840 & 17.2337 \\
kp=0.1   & \textbf{35.2599} \textcolor{ForestGreen}{(+35.19\%)} & \underline{26.0809} & 20.8388 & 14.7310 & 17.6324 \\
\midrule
\multicolumn{6}{c}{\textbf{SNR (kp=0.05, $\Sigma_{\delta}$=0.8)}}\\
\midrule
SNR=0  & \textbf{31.3326} \textcolor{ForestGreen}{(+35.18\%)} & \underline{23.1789} & 16.3286 & 9.8380 & 10.3833 \\
SNR=1  & \textbf{31.3401} \textcolor{ForestGreen}{(+34.32\%)} & \underline{23.3322} & 17.2021 & 11.1680 & 11.6925 \\
SNR=5  & \textbf{31.3330} \textcolor{ForestGreen}{(+33.28\%)} & \underline{23.5085} & 21.4634 & 16.5840 & 17.2337 \\
SNR=10 & \textbf{31.3377} \textcolor{ForestGreen}{(+29.11\%)} & 23.5789 & 22.6505 & 21.9962 & \underline{24.2726} \\
\midrule
\multicolumn{6}{c}{\textbf{$\Sigma_{\delta}$ (kp=0.05, SNR=5)}}\\
\midrule
$\Sigma_{\delta}$=0.0 & \textbf{35.6778} \textcolor{ForestGreen}{(+52.35\%)} & \underline{23.4182} & 21.3606 & 16.5753 & 17.2537 \\
$\Sigma_{\delta}$=0.8 & \textbf{31.3414} \textcolor{ForestGreen}{(+32.89\%)} & \underline{23.5850} & 21.5808 & 16.5840 & 17.2337 \\
$\Sigma_{\delta}$=1.6 & \textbf{30.2325} \textcolor{ForestGreen}{(+30.26\%)} & \underline{23.2091} & 21.3757 & 16.5210 & 17.1866 \\
$\Sigma_{\delta}$=2.4 & \textbf{29.3789} \textcolor{ForestGreen}{(+25.04\%)} & \underline{23.4959} & 21.3979 & 16.3748 & 17.2023 \\
$\Sigma_{\delta}$=3.2 & \textbf{29.1051} \textcolor{ForestGreen}{(+24.43\%)} & \underline{23.3902} & 21.1048 & 16.3402 & 17.2325 \\
\bottomrule
\end{tabular}
\end{adjustbox}
\end{minipage}
\hfill
\begin{minipage}{0.49\textwidth}
\centering
\caption{SSIM comparison under different \textbf{Sampling Rate}, \textbf{SNR}, and \textbf{$\Sigma_{\delta}$} settings (higher is better). The best result in each row is highlighted in bold, the second-best is underlined, and the relative gain is computed as $(\text{best}-\text{second})/\text{second}$.}
\label{tab:ssim_all}
\begin{adjustbox}{max width=\textwidth}
\begin{tabular}{c|ccccc}
\toprule
Setting & RadioDiff-Inv2 (Ours) & RadioDiff-Inv1 & Kriging & RadioUNet & RME-GAN \\
\midrule
\multicolumn{6}{c}{\textbf{Sampling Rate (SNR=5, $\Sigma_{\delta}$=0.8)}}\\
\midrule
kp=0.005 & \textbf{0.8817} \textcolor{ForestGreen}{(+12.95\%)} & 0.5405 & 0.7262 & 0.6431 & \underline{0.7806} \\
kp=0.01  & \textbf{0.8982} \textcolor{ForestGreen}{(+19.94\%)} & 0.5524 & 0.6670 & 0.6045 & \underline{0.7489} \\
kp=0.05  & \textbf{0.9250} \textcolor{ForestGreen}{(+17.54\%)} & \underline{0.7869} & 0.5425 & 0.4898 & 0.7210 \\
kp=0.1   & \textbf{0.9481} \textcolor{ForestGreen}{(+10.77\%)} & \underline{0.8559} & 0.4407 & 0.4112 & 0.7228 \\
\midrule
\multicolumn{6}{c}{\textbf{SNR (kp=0.05, $\Sigma_{\delta}$=0.8)}}\\
\midrule
SNR=0  & \textbf{0.9250} \textcolor{ForestGreen}{(+21.32\%)} & \underline{0.7624} & 0.4408 & 0.2778 & 0.5123 \\
SNR=1  & \textbf{0.9249} \textcolor{ForestGreen}{(+19.08\%)} & \underline{0.7767} & 0.4573 & 0.3151 & 0.5557 \\
SNR=5  & \textbf{0.9250} \textcolor{ForestGreen}{(+17.72\%)} & \underline{0.7857} & 0.5424 & 0.4898 & 0.7210 \\
SNR=10 & \textbf{0.9249} \textcolor{ForestGreen}{(+7.96\%)} & 0.7897 & 0.6157 & 0.6677 & \underline{0.8567} \\
\midrule
\multicolumn{6}{c}{\textbf{$\Sigma_{\delta}$ (kp=0.05, SNR=5)}}\\
\midrule
$\Sigma_{\delta}$=0.0 & \textbf{0.9628} \textcolor{ForestGreen}{(+22.60\%)} & \underline{0.7853} & 0.5420 & 0.4939 & 0.7239 \\
$\Sigma_{\delta}$=0.8 & \textbf{0.9250} \textcolor{ForestGreen}{(+17.26\%)} & \underline{0.7889} & 0.5473 & 0.4898 & 0.7210 \\
$\Sigma_{\delta}$=1.6 & \textbf{0.9198} \textcolor{ForestGreen}{(+17.67\%)} & \underline{0.7817} & 0.5436 & 0.4855 & 0.7199 \\
$\Sigma_{\delta}$=2.4 & \textbf{0.9126} \textcolor{ForestGreen}{(+15.81\%)} & \underline{0.7880} & 0.5444 & 0.4787 & 0.7196 \\
$\Sigma_{\delta}$=3.2 & \textbf{0.9114} \textcolor{ForestGreen}{(+16.46\%)} & \underline{0.7826} & 0.5410 & 0.4748 & 0.7186 \\
\bottomrule
\end{tabular}
\end{adjustbox}
\end{minipage}
\end{table*}

\begin{table*}[!t]
\centering
\begin{minipage}{0.49\textwidth}
\centering
\caption{LPIPS comparison under different \textbf{Sampling Rate}, \textbf{SNR}, and \textbf{$\Sigma_{\delta}$} settings (lower is better). The best result in each row is highlighted in bold, the second-best is underlined, and the relative gain is computed as $(\text{second}-\text{best})/\text{second}$.}
\label{tab:lpips_all}
\begin{adjustbox}{max width=\textwidth}
\begin{tabular}{c|ccccc}
\toprule
Setting & RadioDiff-Inv2 (Ours) & RadioDiff-Inv1 & Kriging & RadioUNet & RME-GAN \\
\midrule
\multicolumn{6}{c}{\textbf{Sampling Rate (SNR=5, $\Sigma_{\delta}$=0.8)}}\\
\midrule
kp=0.005 & \textbf{0.0613} \textcolor{blue}{(-57.13\%)} & 0.3695 & 0.1941 & 0.2762 & \underline{0.1430} \\
kp=0.01  & \textbf{0.0477} \textcolor{blue}{(-69.68\%)} & 0.3451 & 0.2514 & 0.3004 & \underline{0.1573} \\
kp=0.05  & \textbf{0.0286} \textcolor{blue}{(-81.37\%)} & \underline{0.1535} & 0.4045 & 0.3782 & 0.1633 \\
kp=0.1   & \textbf{0.0185} \textcolor{blue}{(-79.30\%)} & \underline{0.0894} & 0.5128 & 0.4479 & 0.1662 \\
\midrule
\multicolumn{6}{c}{\textbf{SNR (kp=0.05, $\Sigma_{\delta}$=0.8)}}\\
\midrule
SNR=0  & \textbf{0.0284} \textcolor{blue}{(-83.32\%)} & \underline{0.1702} & 0.4631 & 0.5670 & 0.2785 \\
SNR=1  & \textbf{0.0285} \textcolor{blue}{(-82.93\%)} & \underline{0.1669} & 0.4642 & 0.5303 & 0.2486 \\
SNR=5  & \textbf{0.0285} \textcolor{blue}{(-82.44\%)} & \underline{0.1623} & 0.4004 & 0.3782 & 0.1633 \\
SNR=10 & \textbf{0.0285} \textcolor{blue}{(-72.62\%)} & 0.1529 & 0.3204 & 0.2535 & \underline{0.1041} \\
\midrule
\multicolumn{6}{c}{\textbf{$\Sigma_{\delta}$ (kp=0.05, SNR=5)}}\\
\midrule
$\Sigma_{\delta}$=0.0 & \textbf{0.0178} \textcolor{blue}{(-88.64\%)} & \underline{0.1566} & 0.4069 & 0.3742 & 0.1625 \\
$\Sigma_{\delta}$=0.8 & \textbf{0.0285} \textcolor{blue}{(-81.78\%)} & \underline{0.1564} & 0.4013 & 0.3782 & 0.1633 \\
$\Sigma_{\delta}$=1.6 & \textbf{0.0316} \textcolor{blue}{(-79.58\%)} & \underline{0.1548} & 0.4015 & 0.3804 & 0.1648 \\
$\Sigma_{\delta}$=2.4 & \textbf{0.0335} \textcolor{blue}{(-77.28\%)} & \underline{0.1474} & 0.4048 & 0.3848 & 0.1653 \\
$\Sigma_{\delta}$=3.2 & \textbf{0.0362} \textcolor{blue}{(-76.87\%)} & \underline{0.1565} & 0.4052 & 0.3881 & 0.1632 \\
\bottomrule
\end{tabular}
\end{adjustbox}
\end{minipage}
\hfill
\begin{minipage}{0.49\textwidth}
\centering
\caption{NMSE comparison under different \textbf{Sampling Rate}, \textbf{SNR}, and \textbf{$\Sigma_{\delta}$} settings (lower is better). The best result in each row is highlighted in bold, the second-best is underlined, and the relative gain is computed as $(\text{second}-\text{best})/\text{second}$.}
\label{tab:nmse_all}
\begin{adjustbox}{max width=\textwidth}
\begin{tabular}{c|ccccc}
\toprule
Setting & RadioDiff-Inv2 (Ours) & RadioDiff-Inv1 & Kriging & RadioUNet & RME-GAN \\
\midrule
\multicolumn{6}{c}{\textbf{Sampling Rate (SNR=5, $\Sigma_{\delta}$=0.8)}}\\
\midrule
kp=0.005 & \textbf{0.066586} \textcolor{blue}{(-34.65\%)} & 0.243266 & 0.104454 & 0.139334 & \underline{0.101889} \\
kp=0.01  & \textbf{0.041929} \textcolor{blue}{(-64.12\%)} & 0.196484 & 0.150055 & 0.140056 & \underline{0.116849} \\
kp=0.05  & \textbf{0.010989} \textcolor{blue}{(-78.93\%)} & \underline{0.052166} & 0.075258 & 0.212081 & 0.140124 \\
kp=0.1   & \textbf{0.004001} \textcolor{blue}{(-86.62\%)} & \underline{0.029898} & 0.085048 & 0.322920 & 0.128131 \\
\midrule
\multicolumn{6}{c}{\textbf{SNR (kp=0.05, $\Sigma_{\delta}$=0.8)}}\\
\midrule
SNR=0  & \textbf{0.010990} \textcolor{blue}{(-80.24\%)} & \underline{0.055630} & 0.236122 & 1.002096 & 0.674556 \\
SNR=1  & \textbf{0.010984} \textcolor{blue}{(-79.85\%)} & \underline{0.054514} & 0.197904 & 0.736565 & 0.499116 \\
SNR=5  & \textbf{0.010989} \textcolor{blue}{(-79.15\%)} & \underline{0.052716} & 0.072380 & 0.212081 & 0.140124 \\
SNR=10 & \textbf{0.010987} \textcolor{blue}{(-60.85\%)} & 0.050245 & 0.066463 & 0.061498 & \underline{0.028066} \\
\midrule
\multicolumn{6}{c}{\textbf{$\Sigma_{\delta}$ (kp=0.05, SNR=5)}}\\
\midrule
$\Sigma_{\delta}$=0.0 & \textbf{0.003916} \textcolor{blue}{(-92.61\%)} & \underline{0.052986} & 0.078509 & 0.211420 & 0.139272 \\
$\Sigma_{\delta}$=0.8 & \textbf{0.010986} \textcolor{blue}{(-78.54\%)} & \underline{0.051193} & 0.070276 & 0.212081 & 0.140124 \\
$\Sigma_{\delta}$=1.6 & \textbf{0.013278} \textcolor{blue}{(-75.90\%)} & \underline{0.055088} & 0.077801 & 0.215149 & 0.141853 \\
$\Sigma_{\delta}$=2.4 & \textbf{0.014795} \textcolor{blue}{(-71.52\%)} & \underline{0.051948} & 0.076699 & 0.222143 & 0.141316 \\
$\Sigma_{\delta}$=3.2 & \textbf{0.016698} \textcolor{blue}{(-69.02\%)} & \underline{0.053901} & 0.090516 & 0.224014 & 0.139965 \\
\bottomrule
\end{tabular}
\end{adjustbox}
\end{minipage}
\end{table*}

\begin{figure*}
\centering
\setlength{\tabcolsep}{2pt}
\renewcommand{\arraystretch}{1.0}

\begin{tabular}{ccccccc}

\includegraphics[width=0.13\textwidth]{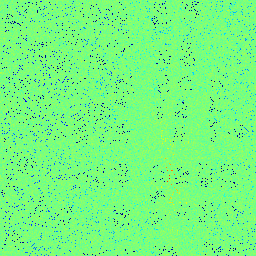} &
\includegraphics[width=0.13\textwidth]{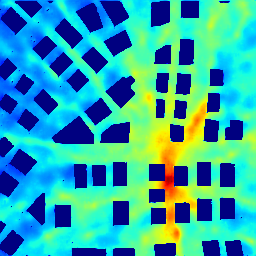} &
\includegraphics[width=0.13\textwidth]{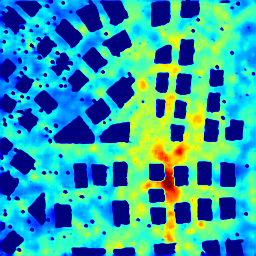} &
\includegraphics[width=0.13\textwidth]{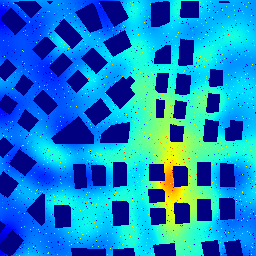} &
\includegraphics[width=0.13\textwidth]{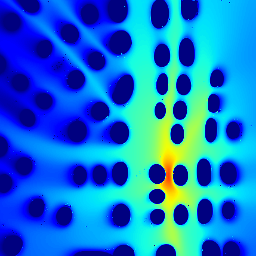} &
\includegraphics[width=0.13\textwidth]{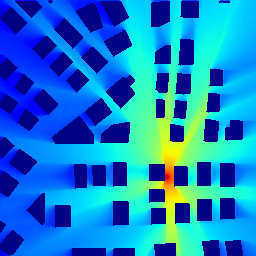} &
\includegraphics[width=0.13\textwidth]{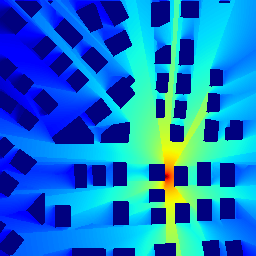} \\
{\scriptsize Samples} & {\scriptsize RME-GAN} & {\scriptsize RadioUNet} & {\scriptsize Kriging} & {\scriptsize RadioDiff-Inv1} & {\scriptsize RadioDiff-Inv2 (Ours)} & {\scriptsize Ground Truth} \\

\includegraphics[width=0.13\textwidth]{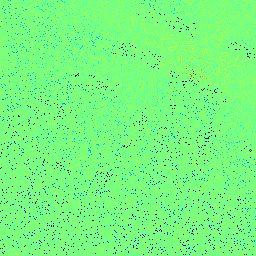} &
\includegraphics[width=0.13\textwidth]{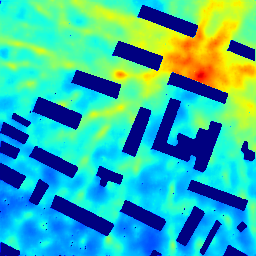} &
\includegraphics[width=0.13\textwidth]{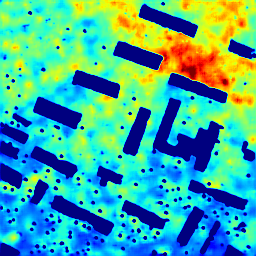} &
\includegraphics[width=0.13\textwidth]{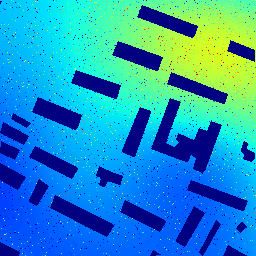} &
\includegraphics[width=0.13\textwidth]{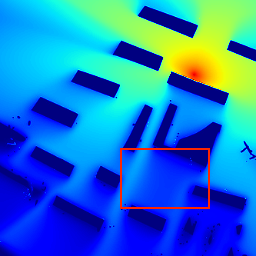} &
\includegraphics[width=0.13\textwidth]{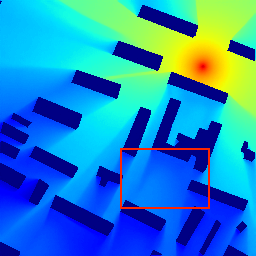} &
\includegraphics[width=0.13\textwidth]{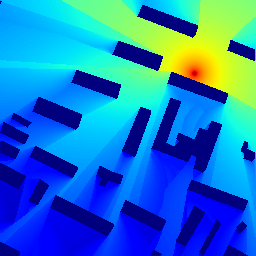} \\
{\scriptsize Samples} & {\scriptsize RME-GAN} & {\scriptsize RadioUNet} & {\scriptsize Kriging} & {\scriptsize RadioDiff-Inv1} & {\scriptsize RadioDiff-Inv2 (Ours)} & {\scriptsize Ground Truth} \\

\includegraphics[width=0.13\textwidth]{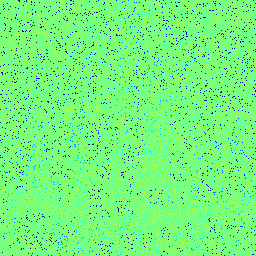} &
\includegraphics[width=0.13\textwidth]{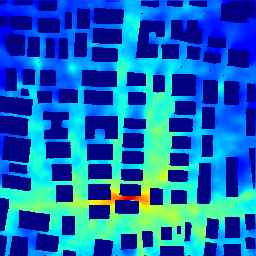} &
\includegraphics[width=0.13\textwidth]{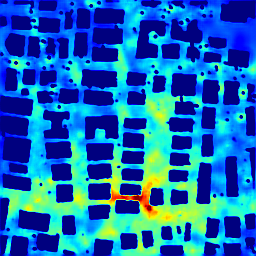} &
\includegraphics[width=0.13\textwidth]{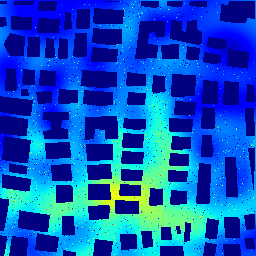} &
\includegraphics[width=0.13\textwidth]{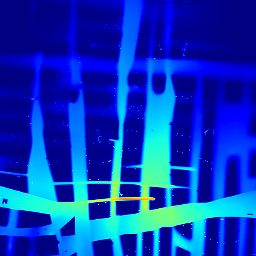} &
\includegraphics[width=0.13\textwidth]{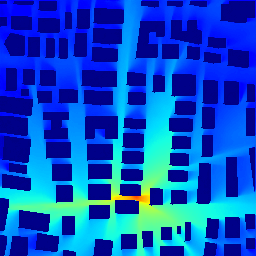} &
\includegraphics[width=0.13\textwidth]{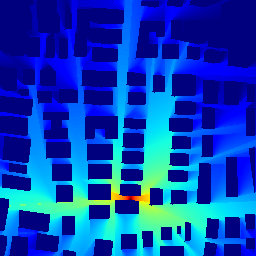} \\
{\scriptsize Samples} & {\scriptsize RME-GAN} & {\scriptsize RadioUNet} & {\scriptsize Kriging} & {\scriptsize RadioDiff-Inv1} & {\scriptsize RadioDiff-Inv2 (Ours)} & {\scriptsize Ground Truth} \\

\includegraphics[width=0.13\textwidth]{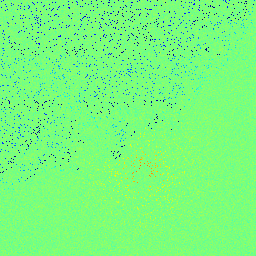} &
\includegraphics[width=0.13\textwidth]{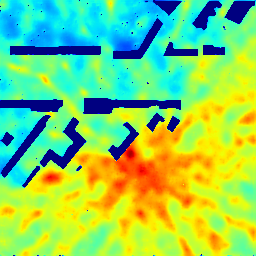} &
\includegraphics[width=0.13\textwidth]{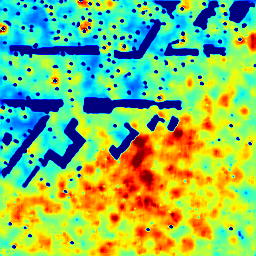} &
\includegraphics[width=0.13\textwidth]{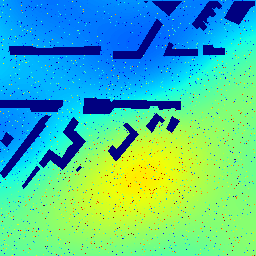} &
\includegraphics[width=0.13\textwidth]{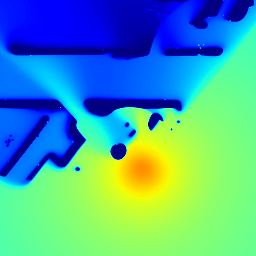} &
\includegraphics[width=0.13\textwidth]{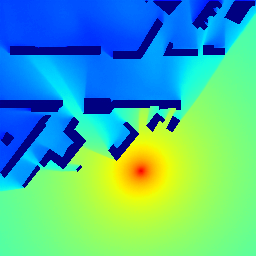} &
\includegraphics[width=0.13\textwidth]{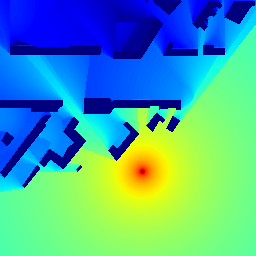} \\
{\scriptsize Samples} & {\scriptsize RME-GAN} & {\scriptsize RadioUNet} & {\scriptsize Kriging} & {\scriptsize RadioDiff-Inv1} & {\scriptsize RadioDiff-Inv2 (Ours)} & {\scriptsize Ground Truth} \\

\end{tabular}

\caption{Visual comparison under \textbf{$\Sigma_{\delta}=0.8$}, $kp=0.05$, and SNR=5.}
\label{fig:vis_sigma08}
\end{figure*}

\subsection{Quantitative Comparison}

Tables~\ref{tab:psnr_all}to \ref{tab:nmse_all} report the results under varying sampling rates with fixed $\mathrm{SNR}=5\,\mathrm{dB}$ and $\Sigma_\delta=0.8$. RadioDiff-Inv2 consistently achieves the best performance across all four metrics and all sampling densities.

In terms of PSNR, RadioDiff-Inv2 surpasses the best competing method by $+18.92\%$ at $kp=0.005$, and the margin grows substantially with increasing sampling density, reaching $+69.20\%$ at $kp=0.1$. This trend indicates that the learned diffusion prior provides increasingly effective guidance as more measurement constraints become available to anchor the inversion. At $kp=0.1$, the proposed method attains $35.26\,\mathrm{dB}$, compared to $20.84\,\mathrm{dB}$ for Kriging—the strongest competing baseline at that point—demonstrating an absolute gain of more than $14\,\mathrm{dB}$.

Consistent improvements are observed in SSIM: the gap rises from $+12.95\%$ at the sparsest setting to $+31.17\%$ at $kp=0.1$, where RadioDiff-Inv2 achieves an SSIM of $0.9481$ against $0.7228$ for RME-GAN. For LPIPS, the proposed method reduces the perceptual distortion by $57.13\%$ to $88.87\%$ relative to the best baseline, with the most dramatic improvement at higher sampling rates. For NMSE, RadioDiff-Inv2 yields a reduction of $34.64\%$ at $kp=0.005$ and $95.26\%$ at $kp=0.1$, reflecting near-exact signal-level reconstruction at moderate-to-high sampling densities.

Notably, the baselines exhibit diminishing or non-monotone improvements as $kp$ increases. Kriging and RadioUNet see their SSIM degrade at higher sampling rates, likely due to noise amplification and overfitting to inaccurate measurement coordinates. In contrast, RadioDiff-Inv2 demonstrates robust and monotone improvement, benefiting from the strong generative prior that constrains the solution to physically plausible RM manifolds.

The SNR ablation ($kp=0.05$, $\Sigma_\delta=0.8$) reveals the most distinctive advantage of RadioDiff-Inv2. As shown in Tables~\ref{tab:psnr_all}--\ref{tab:nmse_all}, the PSNR of RadioDiff-Inv2 remains nearly constant across all SNR levels, hovering around $31.33\,\mathrm{dB}$ from $\mathrm{SNR}=0$ to $\mathrm{SNR}=10$. This near-constant behavior contrasts sharply with the baselines, which degrade rapidly under low SNR.

At $\mathrm{SNR}=0\,\mathrm{dB}$, Kriging, RadioUNet, and RME-GAN achieve PSNR values of $16.33$, $9.84$, and $10.38\,\mathrm{dB}$, respectively—degradations of more than $6\,\mathrm{dB}$ compared to their $\mathrm{SNR}=10\,\mathrm{dB}$ performance. RadioDiff-Inv2, by contrast, maintains $31.33\,\mathrm{dB}$ PSNR at $\mathrm{SNR}=0$, yielding a gain of $+91.89\%$ over the best baseline. The SSIM improvement at $\mathrm{SNR}=0$ reaches $+80.56\%$ ($0.9250$ vs.\ $0.5123$), and the NMSE is reduced by $95.36\%$.

This robustness under low SNR stems from the implicit regularization effect of the diffusion prior. The learned generative model constrains reconstructions to lie on the manifold of physically realistic RMs, effectively decoupling reconstruction quality from the severity of measurement noise. When noise is high, the measurement data carry little useful information, and the prior becomes the dominant factor determining the reconstructed field. For conventional methods that rely solely on interpolation or regression, high noise corrupts the anchor points, leading to severe degradation. For RadioDiff-Inv2, the seed-space optimization adapts to the available signal while the generative prior prevents the estimate from diverging into noise-dominated solutions.

Tables~\ref{tab:psnr_all} to \ref{tab:nmse_all} report the performance under varying drift scales ($kp=0.05$, $\mathrm{SNR}=5\,\mathrm{dB}$) with $\Sigma_\delta$ ranging from $0.0$ (no drift) to $3.2$ (severe drift).

At $\Sigma_\delta=0.0$, RadioDiff-Inv2 achieves $35.68\,\mathrm{dB}$ PSNR and $0.9628$ SSIM, representing a $+67.03\%$ and $+33.00\%$ improvement over the best baseline, respectively. As drift increases to $\Sigma_\delta=3.2$, the PSNR of RadioDiff-Inv2 decreases moderately from $35.68$ to $29.11\,\mathrm{dB}$—a relative drop of only $6.6\,\mathrm{dB}$—while the baseline methods show similar absolute performance but with much larger relative degradation. The PSNR advantage over the best competitor remains above $+37.30\%$ across all drift levels. Similarly, the SSIM of RadioDiff-Inv2 degrades gracefully from $0.9628$ to $0.9114$ as $\Sigma_\delta$ grows, while competitive SSIM values for Kriging and RadioUNet remain below $0.55$ and $0.49$, respectively.

This graceful degradation under increasing drift is attributed to the drift-aware Gaussian resampling operator, which approximates the drift-marginalized likelihood and propagates gradients through the coordinate uncertainty. As a result, the inversion objective remains informative even when individual measurements are severely misregistered, whereas methods that query the field at fixed reported coordinates suffer compounding bias as drift magnitude grows.

\subsection{Visual Comparison}

Figures~\ref{fig:vis_kp005}--\ref{fig:vis_sigma08} provide qualitative comparisons across the three experimental conditions.

Under the default setting ($kp=0.05$, $\mathrm{SNR}=5$, $\Sigma_\delta=0.8$, Fig.~\ref{fig:vis_kp005}), the sparse measurement maps (leftmost column) contain only scattered point observations, making the reconstruction problem severely underdetermined. RME-GAN produces reconstructions with sharp building-level structure but introduces spurious hotspots and color artifacts that are not present in the ground truth. RadioUNet yields blurry outputs with incorrect signal-strength gradients near building edges and occluded regions. Kriging produces overly smooth estimates that fail to recover fine propagation structures. In contrast, RadioDiff-Inv2 accurately reconstructs both the large-scale shadowing patterns and the fine spatial transitions near building boundaries, closely matching the ground truth across all four test samples. Under high SNR ($\mathrm{SNR}=10$, Fig.~\ref{fig:vis_snr10}), the baselines show modest improvements in large-scale field recovery, yet still exhibit local artifacts. RadioDiff-Inv2 maintains precise reconstruction quality and continues to correctly capture sharp shadow edges and propagation corridors, demonstrating that its advantage persists even when measurement quality is favorable. Under Gaussian location drift ($\Sigma_\delta=0.8$, Fig.~\ref{fig:vis_sigma08}), the effect of coordinate uncertainty becomes apparent: RME-GAN and RadioUNet exhibit spatial misalignment between predicted and ground-truth propagation patterns, with incorrect field orientation in several samples. Kriging produces blurred outputs that average over incorrect measurement positions. RadioDiff-Inv2 exhibits no visible misalignment, confirming that the drift-marginalized inversion effectively compensates for coordinate uncertainty and yields reconstructions that are geometrically consistent with the true radio propagation field. Overall, the visual comparisons corroborate the quantitative results and highlight the qualitative superiority of RadioDiff-Inv2 in recovering fine spatial structures while suppressing noise-induced and drift-induced artifacts.

\section{Conclusion}
In this paper, we have proposed RadioDiff-Inv2, a differentiable diffusion inversion framework that addresses the joint challenge of sparse noisy measurements and Gaussian location drift in radio map estimation by formulating a drift-marginalized data-fidelity objective and optimizing a latent noise seed through a deterministic probability-flow ODE, thereby unifying the learned generative prior with a physically consistent, drift-aware measurement operator. Extensive experiments on the RadioMapSeer dataset have demonstrated that RadioDiff-Inv2 consistently outperforms Kriging, RadioUNet, and RME-GAN across all evaluated conditions, achieving up to $14\,\mathrm{dB}$ PSNR gain and over $95\%$ NMSE reduction relative to the best competing baseline, with near-constant reconstruction fidelity across SNR levels from $0$ to $10\,\mathrm{dB}$. The proposed framework offers a practical path toward reliable environment-aware intelligence in 6G networks, where accurate radio maps reconstructed from privacy-constrained crowdsourced feedback can directly enhance coverage planning, interference coordination, and mobility-aware resource management. Future work will explore extending the drift-aware inversion to three-dimensional radio maps and integrating physics-informed diffusion priors to further improve reconstruction accuracy in complex multi-path propagation environments.

\bibliography{ref}

@article{ye2018rmapcs,
  title={RMapCS: Radio map construction from crowdsourced samples for indoor localization},
  author={Ye, Yanzhen and Wang, Bang},
  journal = {Ieee Access},
  volume={6},
  pages={24224--24238},
  year={2018},
  publisher={IEEE}
}

@article{zeng2021toward,
  title={Toward environment-aware {6G} communications via channel knowledge map},
  author={Zeng, Yong and Xu, Xiaoli},
  journal = {Ieee Wireless Commun.},
  volume={28},
  number={3},
  pages={84--91},
  year={2021}
}

@inproceedings{yapar2025sampling,
  title={The sampling-assisted pathloss radio map prediction competition},
  author={Yapar, {\c{C}}a{\u{g}}kan and Bakirtzis, Stefanos and Lutu, Andra and Wassell, Ian and Zhang, Jie and Caire, Giuseppe},
  booktitle={2025 IEEE 35th International Workshop on Machine Learning for Signal Processing (MLSP)},
  pages={1--6},
  year={2025},
  organization={IEEE}
}

@article{wang2025radiodiff,
  title={RadioDiff-inverse: Diffusion enhanced Bayesian inverse estimation for ISAC radio map construction},
  author={Wang, Xiucheng and Fang, Zhongsheng and Cheng, Nan and Sun, Ruijin and Li, Zan and others},
  journal = {Arxiv Preprint Arxiv:2504.14298},
  year={2025}
}

@inproceedings{raivio2003analysis,
  title={Analysis of mobile radio access network using the self-organizing map},
  author={Raivio, Kimmo and Simula, Olli and Laiho, Jaana and Lehtimaki, P},
  booktitle={IFIP/IEEE Eighth International Symposium on Integrated Network Management, 2003.},
  pages={439--451},
  year={2003},
  organization={IEEE}
}

@ARTICLE{11282987,
  author={Wang, Xiucheng and Zheng, Peilin and Jia, Honggang and Cheng, Nan and Sun, Ruijin and Zhou, Conghao and Shen, Xuemin},
  journal = {IEEE Trans. Cogn. Commun. Netw.}, 
  title={RadioDiff-Flux: Efficient Radio Map Construction via Generative Denoise Diffusion Model Trajectory Midpoint Reuse}, 
  year={2026},
  volume={12},
  number={},
  pages={4882-4895}}

@inproceedings{chung2023parallel,
  title={Parallel diffusion models of operator and image for blind inverse problems},
  author={Chung, Hyungjin and Kim, Jeongsol and Kim, Sehui and Ye, Jong Chul},
  booktitle={Proceedings of the IEEE/CVF conference on computer vision and pattern recognition},
  pages={6059--6069},
  year={2023}
}

@article{romero2024theoretical,
  title={Theoretical analysis of the radio map estimation problem},
  author={Romero, Daniel and Ha, Tien Ngoc and Shrestha, Raju and Franceschetti, Massimo},
  journal = {IEEE Trans. Wireless Commun.},
  volume={23},
  number={10},
  pages={13722--13737},
  year={2024},
  publisher={IEEE}
}

@ARTICLE{11083758,
  author={Wang, Xiucheng and Zhang, Qiming and Cheng, Nan and Chen, Junting and Zhang, Zezhong and Li, Zan and Cui, Shuguang and Shen, Xuemin},
  journal = {IEEE Trans. Netw. Sci. Eng.}, 
  title={RadioDiff-3D: A 3D× 3D Radio Map Dataset and Generative Diffusion Based Benchmark for 6G Environment-Aware Communication}, 
  year={2026},
  volume={13},
  number={},
  pages={3773-3789}}

@ARTICLE{11278649,
  author={Wang, Xiucheng and Zhang, Qiming and Cheng, Nan and Sun, Ruijin and Li, Zan and Cui, Shuguang and Shen, Xuemin},
  journal = {IEEE J. Sel. Areas Commun.}, 
  title={RadioDiff-k2: Helmholtz Equation Informed Generative Diffusion Model for Multi-Path Aware Radio Map Construction}, 
  year={2026},
  volume={44},
  number={},
  pages={2318-2333}}

@article{sato2017kriging,
  title={Kriging-based interference power constraint: Integrated design of the radio environment map and transmission power},
  author={Sato, Koya and Fujii, Takeo},
  journal = {IEEE Trans. Cogn. Commun. Netw.},
  volume={3},
  number={1},
  pages={13--25},
  year={2017},
  publisher={IEEE}
}

@article{slijepcevic2002location,
  title={Location errors in wireless embedded sensor networks: sources, models, and effects on applications},
  author={Slijepcevic, Sasha and Megerian, Seapahn and Potkonjak, Miodrag},
  journal = {Acm Sigmobile Mobile Comput. Commun. Review},
  volume={6},
  number={3},
  pages={67--78},
  year={2002},
  publisher={ACM New York, NY, USA}
}

@article{chung2022improving,
  title={Improving diffusion models for inverse problems using manifold constraints},
  author={Chung, Hyungjin and Sim, Byeongsu and Ryu, Dohoon and Ye, Jong Chul},
  journal = {Advances Neural Inf. Process. Syst.},
  volume={35},
  pages={25683--25696},
  year={2022}
}

@inproceedings{le2003mobile,
  title={Mobile location estimator with NLOS mitigation using Kalman filtering},
  author={Le, Bao Long and Ahmed, Kazi and Tsuji, Hiroyuki},
  booktitle={2003 IEEE Wireless Communications and Networking, 2003. WCNC 2003.},
  volume={3},
  pages={1969--1973},
  year={2003},
  organization={IEEE}
}

@inproceedings{venkatraman2002location,
  title={Location using LOS range estimation in NLOS environments},
  author={Venkatraman, Saipradeep and Caffery, James and You, H-R},
  booktitle={Vehicular Technology Conference. IEEE 55th Vehicular Technology Conference. VTC Spring 2002 (Cat. No. 02CH37367)},
  volume={2},
  pages={856--860},
  year={2002},
  organization={IEEE}
}

@inproceedings{yang2013freeloc,
  title={FreeLoc: Calibration-free crowdsourced indoor localization},
  author={Yang, Sungwon and Dessai, Pralav and Verma, Mansi and Gerla, Mario},
  booktitle={2013 Proceedings IEEE INFOCOM},
  pages={2481--2489},
  year={2013},
  organization={IEEE}
}

@article{zhang2020radio,
  title={Radio map-based 3D path planning for cellular-connected UAV},
  author={Zhang, Shuowen and Zhang, Rui},
  journal = {IEEE Trans. Wireless Commun.},
  volume={20},
  number={3},
  pages={1975--1989},
  year={2020},
  publisher={IEEE}
}

@inproceedings{song2020denoising,
  title={Denoising diffusion implicit models},
  author={Song, Jiaming and Meng, Chenlin and Ermon, Stefano},
  booktitle={Proceedings of the 2020 International Conference on Learning Representations (ICLR)},
  pages = {1-12},
  year={2020}
}

@article{6g,
    author = {Cheng, Nan and Chen, Fangjiong and Chen, Wen and Cheng, Zhimi and Yang, Qinghai and Li, Changle and Shen, Xuemin},
    title =  {{6G} omni-scenario on-demand services provisioning: vision, technology and prospect(in Chinese)},
    journal = {Sci Sin Inform},
    year = {2024},
    volume={54},
    pages={1025--1054,}
}

@article{ho2020denoising,
  title={Denoising diffusion probabilistic models},
  author={Ho, Jonathan and Jain, Ajay and Abbeel, Pieter},
  journal = {Advances Neural Inf. Process. Syst.},
  volume={33},
  pages={6840--6851},
  year={2020}
}

@article{shen2023toward,
  title={Toward immersive communications in {6G}},
  author={Shen, Xuemin and Gao, Jie and Li, Mushu and Zhou, Conghao and Hu, Shisheng and He, Mingcheng and Zhuang, Weihua},
  journal = {Frontiers Comput. Science},
  volume={4},
  pages={1068478},
  year={2023},
  publisher={Frontiers Media SA}
}

@article{zhang2023rme,
  title={{RME}-{GAN}: A learning framework for radio map estimation based on conditional generative adversarial network},
  author={Zhang, Songyang and Wijesinghe, Achintha and Ding, Zhi},
  journal = {{IEEE} Internet Things J.},
  year={2023},
  volume={10},
  number={20},
  pages={18016-18027},
  publisher={IEEE}
}

@article{deschamps1972ray,
  title={Ray techniques in electromagnetics},
  author={Deschamps, Georges A},
  journal = {Proc. {IEEE}},
  volume={60},
  number={9},
  pages={1022--1035},
  year={1972},
  publisher={IEEE}
}

@article{wang2024tutorial,
  title={A tutorial on extremely large-scale {MIMO} for {6G}: Fundamentals, signal processing, and applications},
  author={Wang, Zhe and Zhang, Jiayi and Du, Hongyang and Niyato, Dusit and Cui, Shuguang and Ai, Bo and Debbah, M{\'e}rouane and Letaief, Khaled B and Poor, H Vincent},
  journal = {{IEEE} Commun. Surveys Tuts.},
  volume={26},
  number={3},
  pages={1560--1605},
  year={2024},
  publisher={IEEE}
}

@article{levie2021radiounet,
  title={{RadioUNet}: Fast radio map estimation with convolutional neural networks},
  author={Levie, Ron and Yapar, {\c{C}}a{\u{g}}kan and Kutyniok, Gitta and Caire, Giuseppe},
  journal = {{IEEE} Trans. Wireless Commun.},
  volume={20},
  number={6},
  pages={4001--4015},
  year={2021},
  publisher={IEEE}
}

@inproceedings{li2025rmtransformer,
  title={RMTransformer: Accurate radio map construction and coverage prediction},
  author={Li, Yuxuan and Zhang, Cheng and Wang, Wen and Huang, Yongming},
  booktitle={2025 IEEE 101st Vehicular Technology Conference (VTC2025-Spring)},
  pages={1--5},
  year={2025},
  organization={IEEE}
}

@article{jia2025radioflow,
  title={RadioFlow: Efficient Radio Map Construction Framework with Flow Matching},
  author={Jia, Haozhe and Chen, Wenshuo and Wang, Xiucheng and Cheng, Nan and Zhang, Hongbo and Yu, Kuimou and Lai, Songning and Jia, Nanjian and Tian, Bowen and Xiao, Hongru and others},
  journal = {Arxiv Preprint Arxiv:2510.09314},
  year={2025}
}

@article{wang2025iradiodiff,
  title={iRadioDiff: Physics-Informed Diffusion Model for Indoor Radio Map Construction and Localization},
  author={Wang, Xiucheng and Yuan, Tingwei and Cao, Yang and Cheng, Nan and Sun, Ruijin and Zhuang, Weihua},
  journal = {Arxiv Preprint Arxiv:2511.20015},
  year={2025}
}

@inproceedings{jia2025physics,
  title={Physics-informed representation alignment for sparse radio-map reconstruction},
  author={Jia, Haozhe and Chen, Wenshuo and Huang, Zhihui and Wang, Lei and Xiao, Hongru and Jia, Nanqian and Wu, Keming and Lai, Songning and Tian, Bowen and Yue, Yutao},
  booktitle={Proceedings of the 33rd ACM International Conference on Multimedia},
  pages={12352--12360},
  year={2025}
}

@article{li2024radiogat,
  title={RadioGAT: A joint model-based and data-driven framework for multi-band radiomap reconstruction via graph attention networks},
  author={Li, Xiaojie and Zhang, Songyang and Li, Hang and Li, Xiaoyang and Xu, Lexi and Xu, Haigao and Mei, Hui and Zhu, Guangxu and Qi, Nan and Xiao, Ming},
  journal = {IEEE Trans. Wireless Commun.},
  volume={23},
  number={11},
  pages={17777--17792},
  year={2024},
  publisher={IEEE}
}

@article{fang2025radioformer,
  title={RadioFormer: A Multiple-Granularity Radio Map Estimation Transformer with 1$\backslash$textpertenthousand Spatial Sampling},
  author={Fang, Zheng and Liu, Kangjun and Chen, Ke and Liu, Qingyu and Zhang, Jianguo and Song, Lingyang and Wang, Yaowei},
  journal = {Arxiv Preprint Arxiv:2504.19161},
  year={2025}
}

@article{deng2025mars,
  title={MARS: Radio Map Super-resolution and Reconstruction Method under Sparse Channel Measurements},
  author={Deng, Chuyun and Liu, Na and Xie, Wei and Xu, Lianming and Wang, Li},
  journal = {Arxiv Preprint Arxiv:2506.04682},
  year={2025}
}

@article{zhou2025tire,
  title={TiRE-GAN: Task-Incentivized Generative Learning for Radiomap Estimation},
  author={Zhou, Yueling and Wijesinghe, Achintha and Ma, Yibo and Zhang, Songyang and Ding, Zhi},
  journal = {IEEE Wireless Commun. Lett.},
  year={2025},
  publisher={IEEE}
}

@article{chaves2023deeprem,
  title={DeepREM: Deep-learning-based radio environment map estimation from sparse measurements},
  author={Chaves-Villota, Andrea and Viteri-Mera, Carlos A},
  journal = {Ieee Access},
  volume={11},
  pages={48697--48714},
  year={2023},
  publisher={IEEE}
}

@article{wang2025radiodiffloc,
  title={RadioDiff-Loc: Diffusion Model Enhanced Scattering Congnition for NLoS Localization with Sparse Radio Map Estimation},
  author={Wang, Xiucheng and Zhang, Qiming and Cheng, Nan},
  journal = {Arxiv Preprint Arxiv:2509.01875},
  year={2025}
}

@article{zhang2022k,
  title={K-nearest neighbors Gaussian process regression for urban radio map reconstruction},
  author={Zhang, Yifang and Wang, Shaowei},
  journal = {IEEE Commun. Lett.},
  volume={26},
  number={12},
  pages={3049--3053},
  year={2022},
  publisher={IEEE}
}

@article{chen2023probability,
  title={The probability flow ode is provably fast},
  author={Chen, Sitan and Chewi, Sinho and Lee, Holden and Li, Yuanzhi and Lu, Jianfeng and Salim, Adil},
  journal = {Advances Neural Inf. Process. Syst.},
  volume={36},
  pages={68552--68575},
  year={2023}
}

@article{wang2024radiodiff,
  title={{RadioDiff}: An Effective Generative Diffusion Model for Sampling-Free Dynamic Radio Map Construction},
  author={Wang, Xiucheng and Tao, Keda and Cheng, Nan and Yin, Zhisheng and Li, Zan and Zhang, Yuan and Shen, Xuemin},
  journal = {Ieee Trans. Cognit. Commun. Networking, Early Access},
  year={2024},
  pages={1-13},
  publisher={IEEE}
}

@article{wang2026tutorial,
  title={A Tutorial on Learning-Based Radio Map Construction: Data, Paradigms, and Physics-Awarenes},
  author={Wang, Xiucheng and Pan, Yuhao and Cheng, Nan},
  journal={arXiv preprint arXiv:2603.17499},
  year={2026}
}
\bibliographystyle{IEEEtran}
\ifCLASSOPTIONcaptionsoff
  \newpage
\fi
\end{document}